\PassOptionsToPackage{dvipdfmx}{graphicx} 
\documentclass[a4paper,fleqn]{cas-sc}

\usepackage{threeparttable}
\usepackage{lineno}
\usepackage[authoryear]{natbib}
\def\tsc#1{\csdef{#1}{\textsc{\lowercase{#1}}\xspace}}
\tsc{WGM}
\tsc{QE}
\begin{document}
\let\WriteBookmarks\relax
\def\floatpagepagefraction{1}
\def\textpagefraction{.001}

\shorttitle{Numerical Simulations of Hypervelocity Micrometeoroid Impacts}    

\shortauthors{R.Hyodo et al.}  

\title [mode = title]{Numerical Simulations of Hypervelocity Micrometeoroid Impacts: Rocky Impactors onto Icy Targets and the Role of Porosity}  

\tnotemark[1] 

\tnotetext[1]{} 

% First author (corresponding)
\author[1,2,3,4]{Ryuki Hyodo}[orcid=0000-0003-4590-0988]
\cormark[1]
\ead{ryuki.h0525@gmail.com}

\author[5]{Shigeru Wakita}[orcid=0000-0002-3161-3454]
\ead{swakita@purdue.edu}

\author[5,6]{Brandon C. Johnson}[orcid=0000-0002-4267-093X]
\ead{bcjohnson@purdue.edu}
%\fnmark[3]

% Affiliations
\affiliation[1]{organization={Earth-Life Science Institute, Institute of Science Tokyo},
            addressline={2-12-1 Ookayama, Meguro-ku},
            city={Tokyo},
%          citysep={}, % Uncomment if no comma needed between city and postcode
            postcode={152-8550},
            %state={},
            country={Japan}}

\affiliation[2]{organization={Universit\'{e} Paris Cit\'{e}, Institut de Physique du Globe de Paris},
            addressline={1 rue Jussieu},
            city={Paris},
            postcode={75005},
            %state={},
            country={France}}

\affiliation[3]{organization={Graduate School of Artificial Intelligence and Science, Rikkyo University},
            addressline={3-34-1 Nishi-Ikebukuro, Toshima-ku},
            city={Tokyo},
            postcode={171-8501},
            %state={},
            country={Japan}}

\affiliation[4]{organization={SpaceData Inc.},
            addressline={1-17-1 Toranomon, Minato-ku},
            city={Tokyo},
            postcode={105-6490},
            %state={},
            country={Japan}}

\affiliation[5]{organization={Department of Earth, Atmospheric, and Planetary Sciences, Purdue University},
            %addressline={},
            city={West Lafayette, IN},
            %postcode={},
            %state={},
            country={USA}}

\affiliation[6]{organization={Department of Physics and Astronomy, Purdue University},
            %addressline={},
            city={West Lafayette, IN},
            %postcode={},
            %state={},
            country={USA}}

\cortext[1]{Corresponding author: Ryuki Hyodo (ryuki.h0525@gmail.com)}

%\fntext[2]{}


\begin{abstract}
In the outer Solar System, for example in the Saturnian system, a planet's strong gravity attracts micrometeoroids and generates hypervelocity impacts on bodies such as rings and satellites. Micrometeoroids are seemingly non-icy, whereas the targets are typically icy, and both the impactor and the target may span a wide range of porosities. In this study, we perform state-of-the-art three-dimensional iSALE simulations of hypervelocity impacts of rocky impactors onto icy targets, varying the impact angle and the porosities of the impactor and target ($\phi_{\rm imp}$ and $\phi_{\rm tar}$). We consider two end-member porosities (0\% and 90\%) for oblique ($45^\circ$) impacts. At an impact velocity of $30\,\mathrm{km\,s^{-1}}$, characteristic of Saturn's rings, we find that the morphology of early-stage crater formation varies significantly with porosity, transitioning from deep-penetration, narrow-channel cavities ($\phi_{\rm imp}=0$, $\phi_{\rm tar}=90\%$) to very shallow craters driven by near-surface vapor blowoff ($\phi_{\rm imp}=90\%$, $\phi_{\rm tar}=0\%$), with intermediate, more hemispherical cavity shapes when the porosities are comparable. Here, we focus on the thermodynamic fate of the impactor, which represents the exogenic material responsible for modifying the target surface. The impactor material is strongly heated and is efficiently vaporized regardless of the porosities of the impactor and target. However, the peak pressure and peak temperature experienced by the impactor vary by nearly an order of magnitude. These results imply that hypervelocity impacts occurring, for example, in Saturn's rings efficiently vaporize exogenic non-icy impactors upon impact, while the subsequent thermodynamic pathways -- such as condensation and chemical evolution -- may differ depending on the thermodynamic conditions. Our results are expected to be applicable to a variety of planetary systems, particularly in the outer Solar System.
\end{abstract}


%\item Hypervelocity micrometeoroid impacts onto icy targets are investigated using the iSALE-3D shock-physics simulation code.
%\item The porosities of both the impactor and the target independently control shock propagation, energy dissipation, and early-stage crater formation morphology.
%\item Porosity contrast between the impactor and target produces distinct impact regimes, ranging from deep penetration into porous targets to near-surface vapor blowoff on consolidated targets.
%\item These results provide a physical basis for understanding thermodynamic processing, vapor generation, and subsequent chemical evolution during exogenic material impacts on diverse planetary surfaces.
%\item Regardless of porosity, rocky micrometeoroid impactors are efficiently vaporized under hypervelocity conditions typical of Saturn's rings.

% Keywords

\maketitle

% Main text
%%%%%%%%%%%%%%%
% Introduction
%%%%%%%%%%%%%%%
\section{Introduction} 
\label{sec_intro}

The Solar System is populated by micrometeoroids originating from asteroids, comets, and Kuiper-belt debris. The porosity of micrometeoroids, characterized by the void fraction $\phi$, is one of the fundamental physical properties reflecting their parent-body formation and subsequent evolution \citep[e.g.,][]{Blu08,Lev18,Gut19,Blu22}. 

Micrometeoroids that undergo significant thermal alteration during atmospheric entry tend to be melted and/or fused into much denser forms, as observed in micrometeorites collected on Earth \citep{Gen08}. Similarly, micrometeorites from asteroidal sources are generally compact, typically exhibiting moderate porosities \citep[$\phi \sim 10$--$35$\%; e.g.,][]{Bab09}.

In contrast, in the outer Solar System, the Edgeworth-Kuiper belt (EKB, $\sim 30$--$50$ au and beyond) is considered the parent region of Kuiper-belt objects, icy planetesimals, and Jupiter-family comets (JFCs), and a principal source of interplanetary dust particles (IDPs), i.e., micrometeoroids. Micrometeoroids originating from the EKB are typically micron- to submillimeter-sized dust grains. Spacecraft observations, theoretical modeling, and laboratory studies indicate that micrometeoroids originated from the outer solar system are highly diverse, ranging from compact solid grains to fluffy aggregates of submicron grains bound by organic materials and exhibit high porosities \citep[e.g.,][]{Kem23}. Analyses of these micrometeoroids indicate typical porosities of $\phi \sim 50$--$90$\%, reaching over 95\% in extreme fractal cases \citep[e.g.,][]{Fly99,Bro06,Blu06,Lev18,Gut19}.

Depending on their heliocentric orbits, these micrometeoroids impact the solid surfaces of planetary bodies—from Mercury to the icy moons of the outer planets—at hypervelocities of $\sim 10$--$100$\,km\,s$^{-1}$ \citep[e.g.,][]{Pok17,Alt19,Hyo21b,Kem23}. Under such conditions, hypervelocity impacts can flash-vaporize the impactor, melt or amorphize a thin layer of the target, and generate a plume of vapor and melt droplets \citep[e.g.,][]{Mel84}. Consequently, planetary surfaces can be progressively modified and physically and chemically evolve over time through cumulative impact-driven melting, vaporization, mixing, and chemical processing.

Previous studies of impact processes have shown that porosity is not merely a passive property of planetary materials but actively influences shock-wave propagation, energy partitioning, crater morphology, ejecta production, and momentum transfer during impacts \citep[e.g.,][]{Dav10,Kra11,Oka13,Lut17,Col19}. In particular, initial porosity strongly affects melt and vapor production because pore collapse performs additional work, leading to enhanced heating. Although these studies were pioneering, they primarily focused on larger-scale impacts and/or target material, and the separate roles of impactor and target properties in micrometeoroid-scale impacts remain unclear. Therefore, our study focuses, at micrometeoroid scales, on (1) the impactor material, which represents an important carrier of exogenic material that can modify the surface composition of the target, and (2) the dependence on the porosities of both the impactor and the target.

Given the broad parameter space governing micrometeoroid impacts—including impact velocity, composition, and porosity—in this study we focus on rocky (non-icy) impactors striking icy targets, a scenario relevant to outer Solar System environments such as Saturnian system \citep[e.g.,][]{Kem23}. We extend previous studies in two key ways. First, we decouple the porosities of the impactor and target, performing simulations in which either body may be porous while the other is non-porous. Second, we investigate the role of impact geometry by comparing vertical ($90^\circ$) and oblique ($45^\circ$) impacts. This simulation suite enables us to quantify how combinations of impactor-target porosities and impact angle affect the peak pressure and peak temperature experienced by the impactor material, and to qualitatively assess their influence on the early-stage crater formation process.

%%%%%%%%%%%%%%%
% Method
%%%%%%%%%%%%%%%
\section{Numerical Methods} 
\label{sec_method}

We performed numerical simulations of hypervelocity impacts using the iSALE-3D shock-physics code, a multi-material, multi-phase, shock physics code based on a finite-difference formulation in cylindrical or cartesian coordinates \citep{Hirt1974, Collins2004, Elbeshausen2009, Elbeshausen2011}. The code solves the conservation equations of mass, momentum, and energy coupled with appropriate constitutive relations and equations of state (EOS) to describe highly compressible, non-linear material behavior under extreme pressures and temperatures. Artificial viscosity is employed to stabilize shock fronts, and an extension scheme allows the computational domain to follow the growth of the excavation flow and ejecta curtain while minimizing spurious boundary reflections \citep{Dav11, collins2011}.

In our case setup within the iSALE-3D code, material thermodynamics were modeled using the Tillotson-type EOS or ANEOS, which provide pressure as a function of density and internal energy and enable the treatment of condensed, partially vaporized, and fully vaporized states \citep{tillotson1962, Thompson1972}. Solid material response was represented by an elastic--plastic strength model with a pressure-dependent yield criterion and damage accumulation \citep{Mel92, Ivanov1997, Collins2004}. Porosity effects were treated using a strain-based compaction model that accounts for irreversible pore collapse and its influence on shock attenuation, heating, and melt production \citep{Wunnemann2006, collins2011}.

Although our study is motivated by impacts in Saturn's rings, the underlying physical insights are expected to be applicable to other planetary environments, particularly in the outer Solar System. Following \citet{Hyo25}, we assume that a spherical impactor with a radius of $r_{\rm imp} = 10~\mu$m strikes a flat target at an impact velocity of $v_{\rm imp} = 30~\mathrm{km\,s^{-1}}$. Two incident angles are considered, $\theta = 45^\circ$ (oblique impact) and $90^\circ$ (vertical impact). To investigate material dependence relevant to impacts on Saturnian ring particles, we consider a dunite impactor striking a flat icy target (Table~\ref{tab:isale}). Two end-member porosity states, 0\% and 90\%, are examined. The 90\% porosity case should be regarded as a highly porous end-member, rather than a representative value for all micrometeoroids or icy target surfaces. ANEOS parameter sets are employed for both water ice \citep[][Table~\ref{tab:isale}]{Turtle2001} and dunite \citep[][Table~\ref{tab:isale}]{Benz1989}. Since dunite exhibits mechanical and thermodynamic properties similar to those of ordinary chondrites \citep{Svetsov2015}, it serves as a proxy for micrometeorites and interplanetary dust particles.

The spatial resolution is defined in terms of the number of cells per impactor radius (CPPR), and we adopt a uniform resolution of 20 CPPR, corresponding to a cell size of $0.5~\mu$m in the high-resolution zone. This resolution is sufficient to resolve the heated mass and the primary ejecta population \citep{Wakita2019, Luo2022}, while remaining computationally feasible for long-duration simulations ($t \gtrsim 50\,t_s$, where $t_s = 2r_{\mathrm{imp}}/(v_{\mathrm{imp}}\sin\theta)$) and large computational domains required to track the trajectory and thermodynamic fate of the impactor. 

The computational domain was divided into a high-resolution zone and an extension zone. In the extension zone, the cell size increases geometrically with distance from the high-resolution boundary by 3\% per cell, up to a maximum of 20 times the base cell size, following the grid extension scheme. The high-resolution region spans $(-7.5\,r_{\mathrm{imp}},\,7.5\,r_{\mathrm{imp}})$ in the horizontal ($x$) direction, $(-6.0\,r_{\mathrm{imp}},\,6.5\,r_{\mathrm{imp}})$ in the vertical ($z$) direction, and $(0.0\,r_{\mathrm{imp}},\,10\,r_{\mathrm{imp}})$ in the depth ($y$) direction. Boundary conditions were set to mainly outflow, freeslip at $y=0$, and noslip at bottom of the vertical direction as appropriate to minimize numerical reflections and artificial confinement of the ejecta plume.

Lagrangian tracer particles were embedded in each computational cell within the high-resolution zone at the beginning of the simulation ($t=0$). The total number of tracer particles is 4,354,260 for CPPR20 and 4,152,288 for CPPR40, respectively. These tracers record the temporal evolution of particle positions, velocities, pressures, and temperatures, enabling the reconstruction of excavation streamlines, ejection velocities, launch angles, and the thermodynamic fate of the impactor material \citep[e.g.,][]{Joh14,Man22}. 

We compared our three-dimensional iSALE simulations with three-dimensional SPH simulations. We also performed 40 CPPR simulations for comparison. We found good agreement in the peak temperature and peak pressure between the two methods (Appendix~\ref{sec:iSALE-vs-SPH}).

The criterion used to evaluate whether the material is vaporized requires some caution. In this study, we use incipient-vaporization pressure thresholds of $\sim 3$\,GPa for $90$\% porous dunite and $\sim 200$\,GPa for non-porous dunite (Appendix~\ref{appendix_Hugoniot}) as conservative indicators of the onset of vaporization. These threshold corresponds to the critical entropy of $3663$\,J\,K$^{-1}$\,kg$^{-1}$ reported by \citet{Pie97}. 
It should be regarded as a conservative pressure-based criterion, because it is derived from a Hugoniot-based estimate and does not include additional heating mechanisms such as shear/plastic deformation heating \citep{Kurosawa2018,Wakita2019} or post-shock thermal exchange with hot target-derived material, as observed in our simulations (Sec.~\ref{sec:Ppeak_Tpeak}). Therefore, material is expected to melt or vaporize even when the experienced peak pressure is significantly lower than this nominal threshold; further details are provided in Appendix~\ref{appendix_Hugoniot}.

\begin{table}
  \centering
  \begin{threeparttable}
    \begin{minipage}{0.9\linewidth} % or \textwidth など
     \caption{iSALE material input parameters.}
      \label{tab:isale}
    \end{minipage}

\begin{tabular}{lcc}
\hline
\hline
Description & Values & Values \\
\hline
Equation of state &  ANEOS & ANEOS \\
Bulk material of impactor or target & dunite & ice \\
Gruneisen parameter & 0.33 & 0.82 \\
Solidus temperature (K) & 1373 & 273 \\
Simon approximation constant A (MPa) & 1520 &  $\mathrm{NA}$ \\
Simon approximation exponent C & 4.05 & $\mathrm{NA}$ \\
Poisson's ratio & 0.25 & 0.33 \\
Thermal softening parameter & 1.1 & 1.2 \\
Strength model & ROCK & ICE \\
Cohesion, damaged (MPa) & 0.01 & 0.01 \\
Cohesion, undamaged (MPa) & 10 & 10 \\
Frictional coefficient, damaged & 0.6 & 0.6 \\
Frictional coefficient, undamaged & 1.2 & 2.0 \\
Strength at infinite pressure (GPa) & 3.5 & 0.11 \\
Damage model & IVANOV & IVANOV \\
Minimum failure strain & $10^{-4}$ & $10^{-4}$ \\
Damage model constant & $10^{-11}$ & $10^{-11}$ \\
Threshold pressure for damage model (MPa) & 300 & 300 \\
Porosity model & WUNNEMA & WUNNEMA \\
Sound speed ratio & 1.0 & 1.0 \\
Rate of porous compaction & 0.98 & 0.98 \\
Elastic volumetric strain threshold & 0.01 & 0.01 \\
Distension at transition from exponential to power-law compaction & 1.0 &1.0 \\
\hline
\end{tabular}

\begin{tablenotes}[flushleft]
\footnotesize
\item[]  The dunite parameters follow \citet{Wakita2017,Wakita2021}, the ice parameters follow \citet{Bray2014,Sil17}, and the porosity-model parameters follow \citet{Joh15,Wakita2017}. Ice does not use the Simon approximation; the pressure dependence of melt temperature is hard coded. 
\end{tablenotes}

\end{threeparttable}
\end{table}

%%%%%%%%%%%%%%%
% Results
%%%%%%%%%%%%%%%
\section{Numerical Results} 
\label{sec_results}

%%%%%%%%%%%
% Figure 1
%%%%%%%%%%%
\begin{figure}
\begin{center}
	\includegraphics[width=1.0\textwidth]{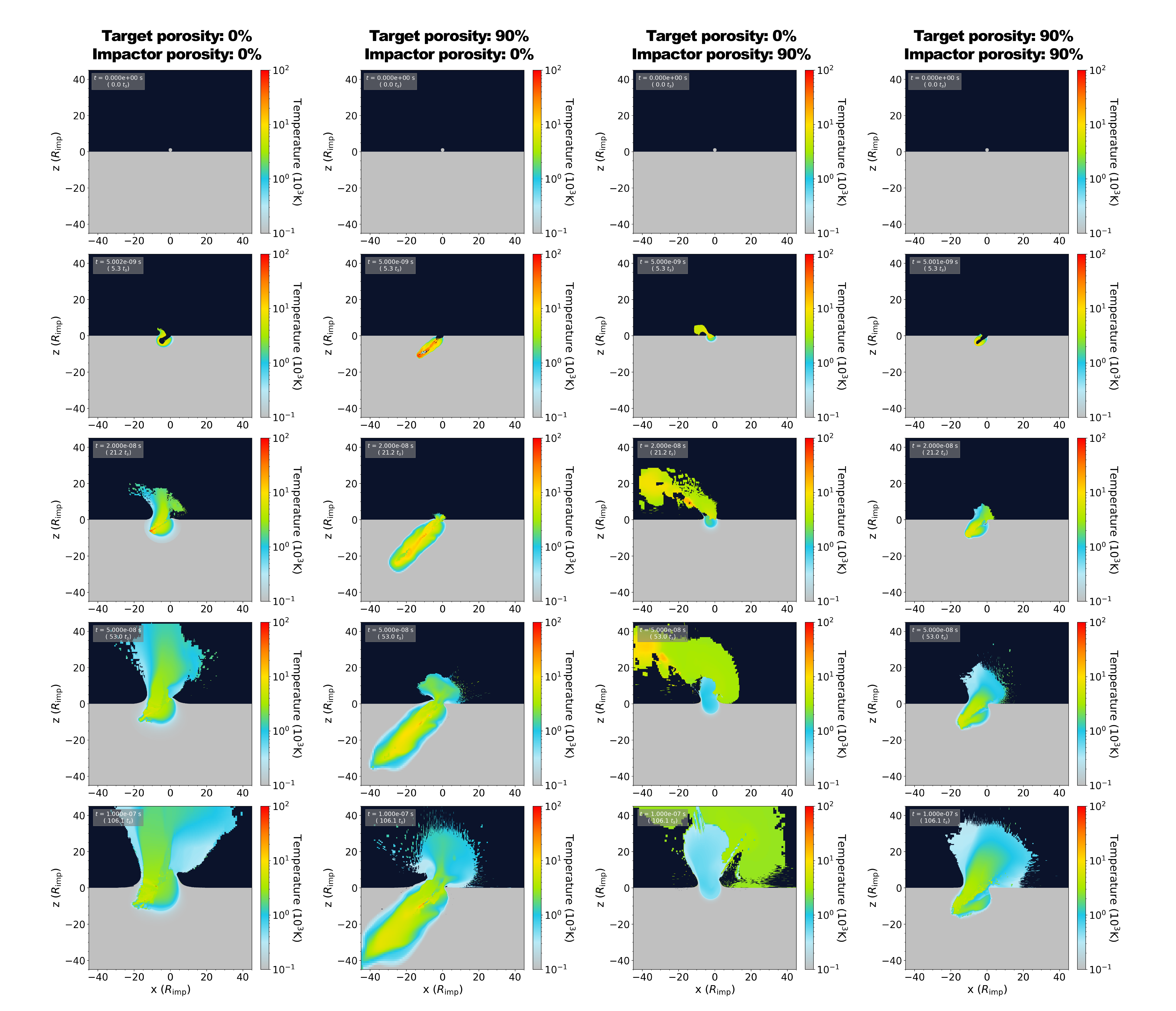}
\caption{Snapshots from iSALE-3D simulations for different combinations of target and impactor porosities, shown for $v_{\rm imp}=30\,\mathrm{km\,s^{-1}}$, $\theta=45^\circ$, $r_{\rm imp}=10\,\mu\mathrm{m}$, and 20 CPPR. Colors indicate temperature.}
\label{Fig_snapshots}
\end{center}
\end{figure}

%-------------------------------------------------------------------------------%
\subsection{Morphological Diversity in Early-Stage Crater Formation}
\label{sec_morphology}

Our simulations do not extend to the final crater stage; therefore, here we focus on the morphology during the early excavation phase. Figure~\ref{Fig_snapshots} demonstrates the overall results of our iSALE-3D simulations for different combinations of target and impactor porosities. The resultant early-stage crater formation varies significantly, ranging from standard crater formation to deep penetration.

Two extreme cases (middle panels in Fig.~\ref{Fig_snapshots}) emerge when there is a stark contrast in porosity between the impactor and the target. When a consolidated impactor ($\phi_{\rm imp}=0$\%) strikes a highly porous target (i.e., $\phi_{\rm tar}=90$\%), significant penetration occurs \citep[see also][]{Lut17}\footnote{
The penetration-like morphology is qualitatively similar to impacts into highly porous snow-like targets modeled by Luther et al. (2017). However, our simulations differ in scale, geometry, materials, and focus: we consider micrometeoroid-scale, three-dimensional oblique impacts of dunite onto water ice at $30~{\rm km~s^{-1}}$ using ANEOS, and examine the thermodynamic fate of the impactor while independently varying both impactor and target porosities.}. In this scenario, heated material ($\gg 1,000$\,K) fills the narrow penetration channel and is subsequently expelled from the entry opening, forming a high-temperature plume. Conversely, when a highly porous impactor (i.e., $\phi_{\rm imp}=90$\%) strikes a consolidated target (i.e., $\phi_{\rm tar}=0$\%), our iSALE simulations show little to no penetration of the impactor into the target and no formation of a deep cavity during the early excavation stage. Instead, the impact-generated hot vapor ($\sim 1,000-10,000$\,K) remains concentrated near the surface and expands rapidly above the impact site.

The left and right panels in Fig.~\ref{Fig_snapshots} show intermediate cases between the two extremes described above. The case with no porosity in either the target or the impactor (left panel; i.e., $\phi_{\rm tar}=0$\% and $\phi_{\rm imp}=0$) has been the primary focus of most of the previous research. The case where both the target and impactor are highly porous is shown in the right panel (i.e., $\phi_{\rm tar}=90$\% and $\phi_{\rm imp}=90$). Comparing these two, the morphology of the early-stage crater formation is more hemispherical and/or rounded, and the vapor expands more rapidly in the zero-porosity case.

 %-------------------------------------------------------------------------------%
 \subsection{Peak Pressure and Peak Temperature Experienced by Impactor Materials}
\label{sec:Ppeak_Tpeak}

%%%%%%%%%%%
% Figure 2
%%%%%%%%%%%
\begin{figure}
\begin{center}
	\includegraphics[width=1.0\textwidth]{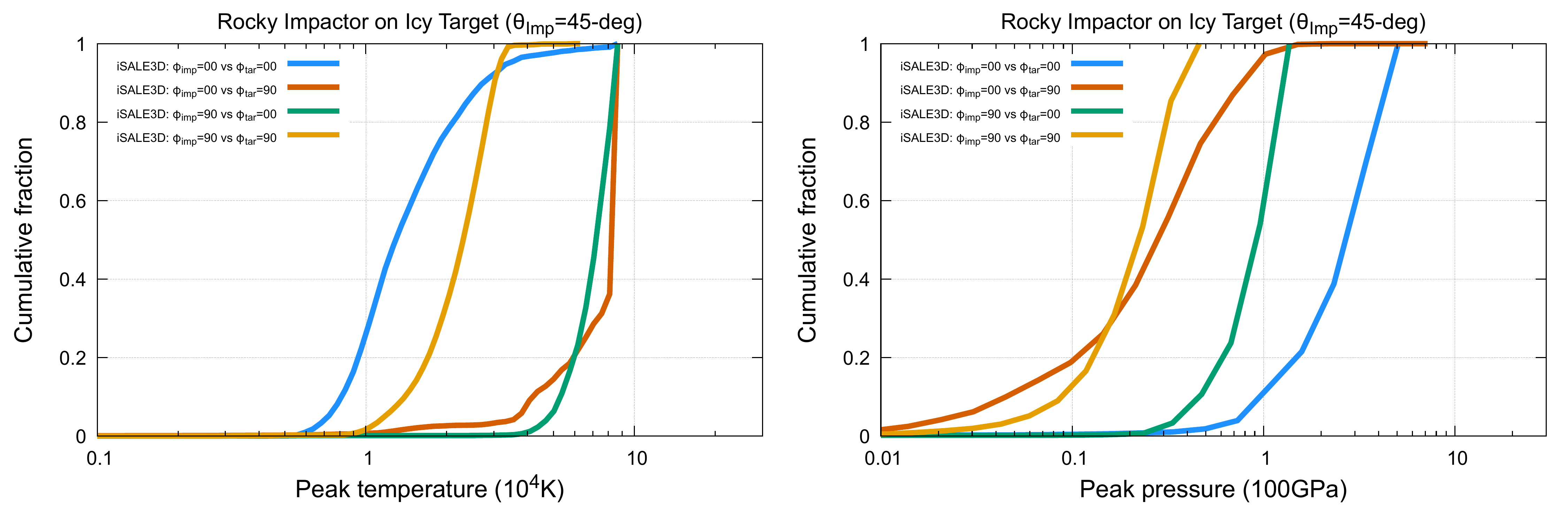}
\caption{Cumulative fraction of the peak temperature (left) and peak pressure (right) for different combinations of impactor and target porosities. Here, target material is not included because the impactor material is the focus of this study. The impact velocity is $30\,\mathrm{km\,s^{-1}}$ and the impact angle is $45^\circ$. Although the peak pressure and temperature vary by nearly an order of magnitude depending on porosity, the impactor material is efficiently vaporized under the hypervelocity conditions considered here; see Sec.~\ref{sec:Ppeak_Tpeak} and Appendix~\ref{appendix_Hugoniot} for further discussion.}
\label{Fig_Ppeak_Tpeak}
\end{center}
\end{figure}

\begin{figure}
\begin{center}
	\includegraphics[width=0.7\textwidth]{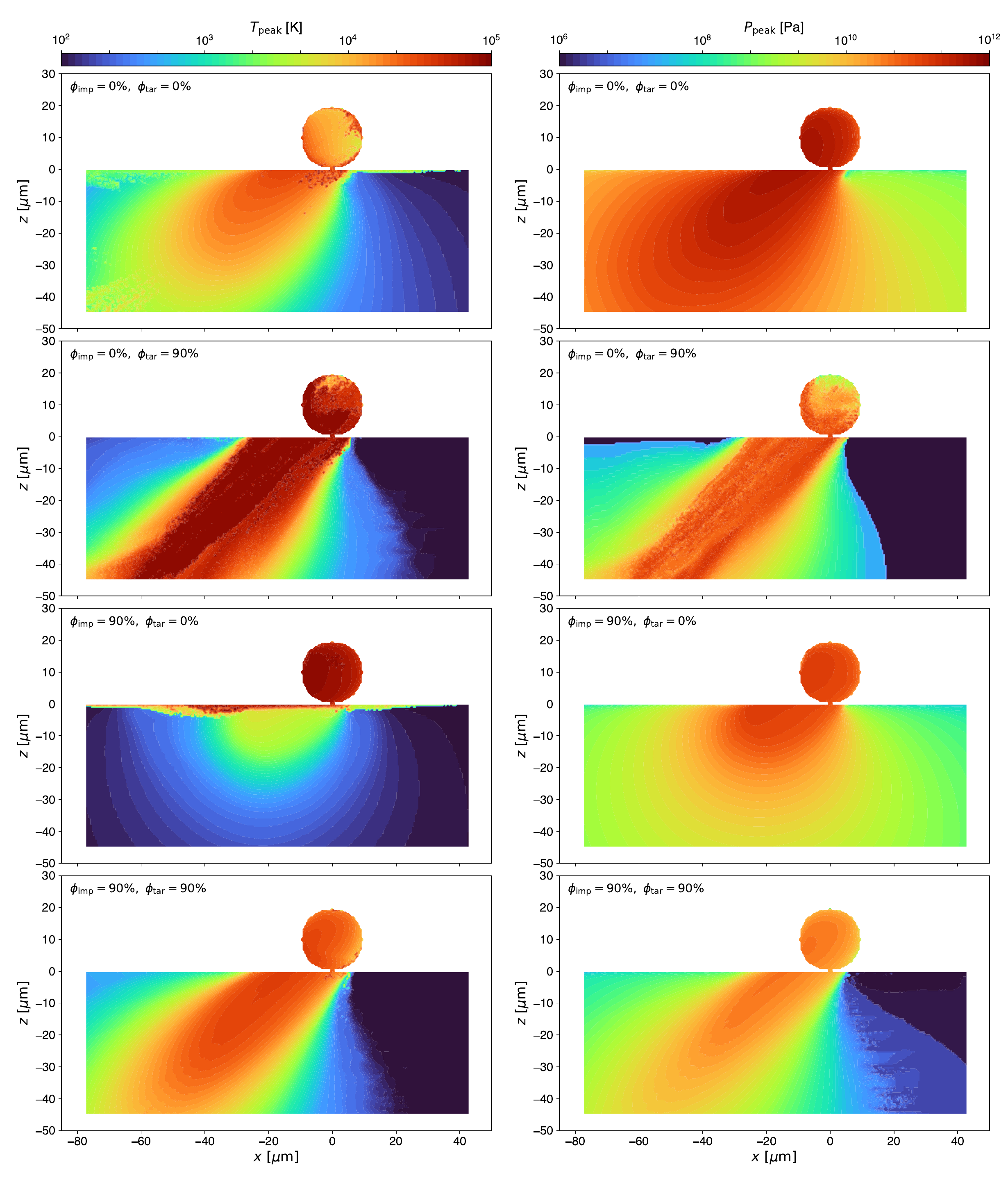}
\caption{Spatial distributions of peak temperature ($T_{\rm peak}$, left column) and peak pressure ($P_{\rm peak}$, right column) mapped onto the initial ($t = 0$) particle positions in the $y \approx 0$ cross-section ($x$--$z$ plane). Each row corresponds to a different combination of impactor and target porosity: $\phi_{\rm imp} = 0\%$, $\phi_{\rm tar} = 0\%$ (first row); $\phi_{\rm imp} = 0\%$, $\phi_{\rm tar} = 90\%$ (second row); $\phi_{\rm imp} = 90\%$, $\phi_{\rm tar} = 0\%$ (third row); and $\phi_{\rm imp} = 90\%$, $\phi_{\rm tar} = 90\%$ (fourth row). Here, the impact angle of $45^\circ$ is used. Note that only the tracer particles are plotted here, and the computational domain is much larger, as shown in Fig.~\ref{Fig_snapshots}.}
\label{Fig_T0snapshot_45deg}
\end{center}
\end{figure}

In this subsection, we examine the peak temperature ($T_{\rm peak}$) and peak pressure ($P_{\rm peak}$) experienced by the impactor material. These quantities were evaluated using the data at the end of each simulation, by which time we confirmed that the results had sufficiently converged. Figure~\ref{Fig_Ppeak_Tpeak} shows the cumulative distribution of the peak temperature (left) and peak pressure (right) for various combinations of target and impactor porosities. These distributions are obtained from tracer particles representing the impactor material (Fig.~\ref{Fig_T0snapshot_45deg}).

Regarding the peak temperature, the smallest temperature increase is observed in the case where both the target and the impactor have no porosity ($\phi_{\rm tar} = 0\%$, $\phi_{\rm imp} = 0\%$; blue line). In contrast, the most significant temperature increases occur in the asymmetric porosity cases, where either the target or the impactor is highly porous while the other is consolidated (green line or red line). When both the target and the impactor are highly porous (orange line; $\phi_{\rm tar} = 90\%$, $\phi_{\rm imp} = 90\%$), the resulting temperature distributions fall between these two extreme cases.

When porosity is present, additional work is performed to compress the pore spaces, leading to a greater temperature increase compared to the non-porous case \citep[see also][]{Dav10,Kra11,Col19}. Consequently, it is expected that when the impactor itself is porous (e.g., the orange and green lines), the impactor material experiences more significant heating during the impact process than in the consolidated impactor cases (e.g., blue line).

However, it may appear counter-intuitive that the case with a consolidated impactor and a highly porous target ($\phi_{\rm imp} = 0\%$ and $\phi_{\rm tar} = 90\%$; red line) also exhibits a substantial temperature increase in the impactor material. This specific case is characterized by significant penetration (middle left panel in Fig.~\ref{Fig_snapshots}). During the penetration process, hot water vapor is expected to be generated from the porous icy target material and fills the resulting penetration channel (Appendix~\ref{appendix_Hugoniot}). Therefore, the observed high temperature of the impactor material in this scenario is likely due to thermal heat exchange between the impactor and the surrounding target-derived hot vapor, rather than the direct result of shock-induced heating within the impactor itself. We will discuss this more in analytical arguments in the following subsection.

Concerning the peak pressure, the distributions exhibit a distinct trend compared to the temperature profiles. The maximum $P_{\rm peak}$ values are generally lower in cases involving high porosity, which can be attributed to significant energy dissipation during the compaction of pore spaces. Consequently, when either both the target and impactor are porous (orange line) or only the target is porous (red line), the peak pressure experienced by the impactor material is minimized. This indicates that the shock delivered to the impactor is effectively alleviated by the porous nature of the target. In contrast, the highest peak pressure is achieved when both the target and the impactor consist of consolidated material (blue line), where shock attenuation is minimal.

Taken together, the complex interplay among shock pressure, temperature increase, porosity-dependent pore-collapse heating, and post-shock thermal exchange suggests that the impactor material is efficiently vaporized under the hypervelocity conditions considered here (see Appendix~\ref{appendix_Hugoniot} for further details).

%---------------------%
% Table 1 
%---------------------%
\begin{table}[t]
\centering
\caption{Material parameters used in the semi-analytical model.}
\label{tab:material_params}
\begin{tabular}{lcccccc}
\hline
Material & $\rho_{s,0}$ (kg\,m$^{-3}$) & $C$ (m\,s$^{-1}$) & $S (-)$ & $C_v$ (J\,kg$^{-1}$\,K$^{-1}$) & $\gamma (-)$ & $\kappa (-)$ \\
\hline
Dunite (impactor) & 3320$^{a}$ & 6600$^{a}$ & 0.90$^{a}$ & 1000 & 0.33$^{c}$ & 0.98$^{e}$ \\
Water ice (target) & 932$^{b}$   & 1700$^{b}$ & 1.44$^{b}$ & 4000 & 0.82$^{d}$ & 0.98$^{e}$ \\
\hline
\end{tabular}
\vspace{2mm}
\begin{minipage}{0.9\linewidth}
\footnotesize
$^{a}$ \cite{Mel84}, $^{b}$ \cite{Ste05}, $^{c}$ \cite{Benz1989}, $^{d}$ Value used in iSALE. $^{e}$ \cite{Wunnemann2006}
\end{minipage}
\end{table}
%---------------------%

%%%%%%%%%%%
% Figure 3
%%%%%%%%%%%

 %-------------------------------------------------------------------------------%
\section{Semi-Analytical Estimation of Peak Pressure and Peak Temperature in High-Velocity Impacts}
\label{sec:semi-analytical}

%%%%%%%%%%%
% Figure 4
%%%%%%%%%%%
\begin{figure}
\begin{center}
	\includegraphics[width=1.0\textwidth]{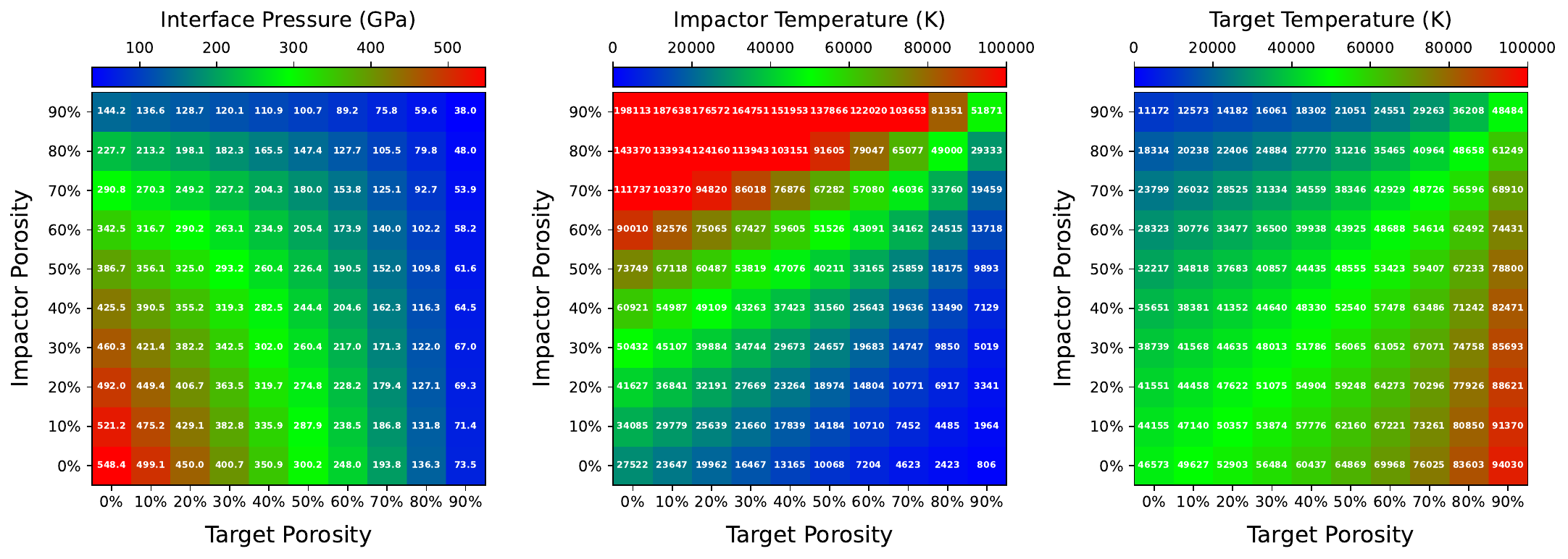}
\caption{Semi-analytical predictions of the peak shock pressure and post-shock temperature as a function of the target and impactor porosities. Left: peak pressure $P_{\rm peak}$ at the interface, which is identical for the target and impactor under the impedance-matching condition. Middle: peak temperature of the impactor material, $T_{\rm peak,imp}$. Right: peak temperature of the target material, $T_{\rm peak,tar}$. Here we consider a reference case with a normal impact at $v_{\rm imp}=30\,\mathrm{km\,s^{-1}}$ and the material parameters listed in Table~\ref{tab:material_params}.}
\label{Fig_analytical_reference}
\end{center}
\end{figure}

%%%%%%%%%%%
% Figure 5
%%%%%%%%%%%
\begin{figure}
\begin{center}
	\includegraphics[width=1.0\textwidth]{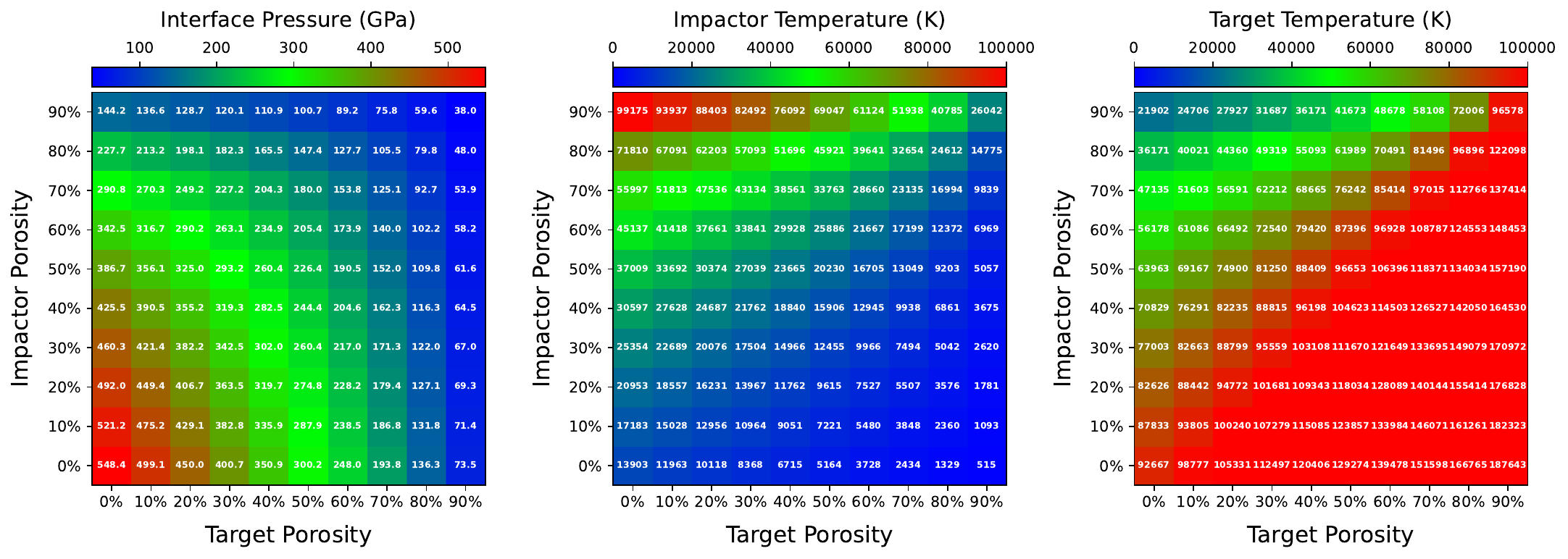}
\caption{Same as Fig.~\ref{Fig_analytical_reference}, except that $C_v$ is set to $2000\,\mathrm{J\,kg^{-1}\,K^{-1}}$ for both the target and the impactor.}
\label{Fig_analytical_CV}
\end{center}
\end{figure}

Here we present a semi-analytical model to estimate the peak shock pressure and peak temperature attained in both the target and the impactor. A full derivation of the model is provided in Appendix~\ref{sec:derivation}.

The model is constructed based on (i) a shock-jump (Hugoniot) description for each material and (ii) an interface (impedance-matching) condition used to determine the peak pressure. The corresponding temperature increase is then estimated by partitioning the shock energy into reversible compression of the solid matrix and irreversible dissipation associated with pore collapse (when porosity is present).

For a given set of material parameters $(\rho_{s,0}, \phi, C, S, \gamma, C_v)$, together with a specified normal impact velocity $v_{\rm imp,nor}$ and compaction parameter $\kappa$, the model predicts $P_{\rm peak}$ and $T_{\rm peak}$. This framework serves as a transparent baseline that complements the fully numerical simulations. Here, $\rho_{s,0}$ denotes the reference (pore-free) solid density, and $\phi$ is the initial porosity. The parameters $C$ and $S$ are the coefficients in the linear Hugoniot relation $U_s = C + S u_p$, where $U_s$ is the shock velocity and $u_p$ is the particle velocity \citep{Mel84}. The parameter $\gamma$ is the Grüneisen parameter, and $C_v$ is the specific heat at constant volume. The material parameters used in this study are summarized in Table~\ref{tab:material_params}.

Figure~\ref{Fig_analytical_reference} shows the predicted $P_{\rm peak}$ and $T_{\rm peak}$ for both the target and the impactor. Unless otherwise noted, we adopt a reference case with $v_{\rm imp,nor}=30\,\mathrm{km\,s^{-1}}$ and the material parameters listed in Table~\ref{tab:material_params}.

Because we assume impedance matching at the interface, the peak pressure (left panel) is identical in the target and the impactor. Overall, increasing either the target porosity or the impactor porosity reduces $P_{\rm peak}$, reflecting enhanced energy dissipation during pore collapse. Accordingly, the largest peak pressure is obtained for non-porous targets and impactors, whereas the smallest peak pressure is obtained when both are highly porous. This dependence on target and impactor porosities is consistent with the results of our direct iSALE-3D simulations (see the right panel of Fig.~\ref{Fig_Ppeak_Tpeak}).

The predicted peak temperature of the impactor material is shown in the middle panel of Fig.~\ref{Fig_analytical_reference}. Increasing the target porosity while keeping the impactor porosity fixed reduces the peak temperature experienced by the impactor. This is because a larger fraction of the impact energy is consumed by compaction of the porous target, thereby lowering the shock heating of the impactor material. In contrast, increasing the impactor porosity while keeping the target porosity fixed increases the peak temperature of the impactor. In this case, additional irreversible dissipation associated with compaction of the impactor's own pore space contributes directly to heating within the impactor material.

Finally, the predicted peak temperature of the target material is shown in the right panel of Fig.~\ref{Fig_analytical_reference}. The trend is opposite to that in the middle panel. Increasing the target porosity while keeping the impactor porosity fixed increases the peak temperature experienced by the target material, whereas increasing the impactor porosity while keeping the target porosity fixed decreases the peak temperature of the target material. The underlying mechanism is the same as discussed above, but here we focus on the thermodynamic response of the target material.

By additionally considering the early-stage crater formation morphology (see Sec.~\ref{sec_morphology}), we can draw further insights into the impact outcome. A particularly instructive limiting case is a non-porous (consolidated) impactor (e.g., $\phi_{\rm imp}=0$) striking a highly porous target (e.g., $\phi_{\rm tar}=90$). In this case, the semi-analytical model (Fig.~\ref{Fig_analytical_reference}) predicts that the impactor is heated only to $\sim 10^3$\,K, whereas the target reaches temperatures of order $\sim 10^5$\,K. The early-stage crater formation morphology (i.e., Fig.~\ref{Fig_snapshots}) further shows that this parameter combination ($\phi_{\rm imp}=0$ and $\phi_{\rm tar}=90$) leads to deep penetration of the impactor into the target. Therefore, even if the impactor is not strongly heated by direct shock compression, heat exchange with the surrounding hot vapor generated from the target can substantially increase the impactor temperature. This is consistent with our direct iSALE-3D simulations (see the red line in the left panel of Fig.~\ref{Fig_Ppeak_Tpeak}).

Variations in material properties can change the results quantitatively, among which the specific heat at constant volume, $C_v$, exerts the strongest influence on the predicted temperature. The sensitivity to other parameters is comparatively weak. In particular, the dependence on $C$, $S$, and $\gamma$ is modest in our parameter range, and these quantities typically vary within a limited range for the materials considered. We adopt $\kappa=0.98$ following standard iSALE implementations \citep{Wunnemann2006}. In the hypervelocity regime examined here, pore collapse is generally complete, and therefore the influence of $\kappa$ on the peak state is minor.

Increasing $C_v$ reduces the temperature rise for a given energy input, $\Delta T = \Delta E/C_v$, and therefore lowers the predicted peak temperature and vice versa (Fig.~\ref{Fig_analytical_CV}). However, the qualitative dependence on porosity remains unchanged. Therefore, this framework is useful for capturing the underlying physics.

The comparison between the semi-analytical model and the iSALE-3D simulations should be interpreted with caution, because the two approaches include different levels of physical complexity. The semi-analytical model reproduces the first-order decrease in peak pressure with increasing impactor or target porosity, because this trend is mainly controlled by impedance matching and pore-collapse dissipation. It also captures the direct increase in impactor temperature with increasing impactor porosity at fixed target porosity, because pore-collapse work is deposited within the impactor material. However, the semi-analytical model does not include the mechanism responsible for the enhanced impactor temperature seen in the iSALE-3D simulation for the case of a non-porous impactor striking a highly porous target (i.e., $\phi_{\rm imp} = 0\%$ and $\phi_{\rm tar} = 90\%$). In this case, the temperature increase is not primarily caused by direct shock heating of the impactor, but by post-shock heat exchange with hot vapor generated from the porous target during deep penetration (see Section~\ref{sec:Ppeak_Tpeak}). This process involves multidimensional penetration, vapor expansion, and material mixing, which are not included in the semi-analytical model. Therefore, the semi-analytical model should be regarded as a baseline estimate for direct shock and pore-collapse heating, while the iSALE-3D simulations additionally capture geometry-dependent post-shock thermal exchange.

%%%%%%%%%%%%%%%
% Discussion
%%%%%%%%%%%%%%%
\section{Discussion}
\label{sec_discussion}

Several limitations of the present study should be noted. These include the limited range of impact velocities, material compositions, and the adopted porosity model with only two end-member values. In reality, micrometeoroid populations exhibit a broad range of velocities, compositions, and internal structures. Moreover, long-term evolutions of crater formation, vapor expansion within and outside the crater, and condensates were not modeled in the present framework. Our iSALE-3D simulations are computationally demanding, and each simulation run in this study required approximately one month on a high-performance computing resources, with limited access to them. In future work, we plan to explore a broader parameter space, including impacts of icy impactors onto rocky targets.

In the present study, we adopted a simplified homogeneous porosity model that represents a continuum approximation; therefore, grain-scale heterogeneity is not resolved. When the size of the impactor becomes comparable to the characteristic scale of the porosity, a more realistic scenario corresponds to a collision between the impactor and an individual constituent grain of the target material. In such cases, the porosity distribution is no longer homogeneous, and the impact geometry may play a more significant role in controlling the impact outcome. Thus, the dependence on the constituent grain size, sizes of impactor and target may play an important role.

The applicability of the adopted $\varepsilon$--$\alpha$ compaction model should be interpreted with caution in the very high-porosity regime. Previous studies have shown that this type of strain-based compaction model can reproduce experimental Hugoniot data reasonably well for highly porous materials up to porosities of approximately 70--80\% \citep[e.g.,][]{collins2011}, and it has also been applied to penetration into highly porous snow-like materials with porosities of order 70\% \citep[e.g.,][]{Lut17}. However, porosities approaching 90\% are less well constrained experimentally and may involve extrapolation of the compaction model. At such high porosities and at micrometeoroid scales, real materials may have aggregate-like or granular structures, and the assumptions of homogeneous porosity and continuum behavior may become questionable when the pore or grain scale is not sufficiently smaller than the impactor size. Future dedicated modeling that resolves grain-scale or aggregate-scale structure will be required to assess these effects more quantitatively.

The EOS uncertainty is also important in the very high-porosity regime. \cite{Kra11} showed that porous-ice Hugoniot calculations using the five-phase EOS of \cite{Ste08} increasingly deviate from experimental data above porosities of approximately 60\%. Although we use ANEOS for water ice, the thermodynamic response of ice at 90\% porosity remains poorly constrained experimentally. Thus, the peak temperatures and vaporization behavior in the 90\% target-porosity cases should be interpreted as qualitative end-member results rather than precise quantitative predictions.

As a separate limitation, because the impactor radius and velocity are fixed in our simulations, a porous impactor also has a smaller bulk mass and kinetic energy; a 90\% porous impactor has only 10\% of the mass of a non-porous impactor with the same radius. Thus, part of the weaker pressure increase or excavation for porous impactors may reflect the smaller impactor mass. However, this case would not be equivalent to a smaller non-porous impactor, because pore collapse, reduced shock impedance, and irreversible compaction heating alter the shock and thermal evolution.

Regarding the applied numerical techniques, we note that the heat exchange in iSALE simulations may be primarily driven by material mixing and numerical diffusion associated with the advection scheme, rather than by explicitly resolved physical thermal conduction. In reality, when strong temperature gradients develop (as in the present penetration regime), thermal equilibration would occur through physical heat conduction and turbulent mixing at small scales. However, iSALE does not directly solve the heat conduction equation; instead, temperature homogenization arises from numerical mixing inherent to the Eulerian advection scheme. In our simulations, the impactor radius is $R_{\rm imp}=10\,\mu{\rm m}$ and the spatial resolution corresponds to CPPR = 20, giving a cell size of $\Delta x = R_{\rm imp}/{\rm CPPR}$.  Accordingly, temperature changes within the mixed region occur on the dynamical/advection timescale (of order $10^{-8}\,{\rm s}$ in our runs; see Fig.~\ref{Fig_snapshots}). For reference, a purely conductive equilibration timescale over one cell width would be $\tau_{\rm cond}\sim \Delta x^2/\alpha_{\rm th}$; adopting $\alpha_{\rm th} \sim10^{-5} - 10^{-6}\,{\rm m^2\,s^{-1}}$ yields $\tau_{\rm cond}\sim 2.5\times10^{-8} - 2.5\times10^{-7}\,{\rm s}$. Thus, the numerical homogenization timescale in iSALE is somewhat shorter than the estimated physical conductive timescale. Nevertheless, both timescales are of the same order of magnitude, and the penetration process itself occurs on a comparable dynamical timescale. Therefore, although the mechanism of temperature equilibration in iSALE is numerical rather than explicitly conductive, the effective thermal exchange rate remains physically plausible for the hypervelocity regime considered here. In this sense, the rapid temperature homogenization seen in iSALE can be interpreted as a numerically mediated representation of physically expected heat exchange during impact-induced mixing, and our results can be regarded as a reasonable approximation to real thermodynamic evolution under such extreme conditions.

In this study, we focus on the impactor material to investigate its thermodynamic fate under hypervelocity impacts ($\sim 30\,\mathrm{km\,s^{-1}}$), motivated in part by micrometeoroid impacts in Saturn's rings. Such micrometeoroid impacts have been considered a source of exogenic material that pollutes Saturn's rings, and the degree of pollution has been used to estimate the age of the rings \citep[e.g.,][]{Zha17a}. Our results indicate that micrometeoroids—i.e., potential polluting material—are vaporized upon impact; therefore, micrometeoroid material does not directly pollute the rings in its original form without undergoing significant thermodynamic and chemical modification \citep{Hyo25}.

Similar hypervelocity impacts are expected to occur on many other planetary bodies beyond Saturn, where such impacts may contribute to the physical and chemical evolution of surface materials. These processes can modify surface structure, composition, and thermodynamic states over time. Therefore, it is important to extend our investigation to a broader range of planetary environments. Although comparable impact velocities may occur across different bodies, exploring a wider range of impact velocities, together with varying combinations of impactor-target composition, porosity, and other material properties, is essential to better constrain impact outcomes under diverse planetary conditions.

The semi-analytical model developed in this study reproduces the overall trends of peak pressure and temperature obtained from the numerical simulations and provides a transparent physical interpretation of how energy is partitioned between reversible compression and irreversible pore-collapse dissipation. While the model captures the first-order dependence on porosity, it does not include geometric effects such as deep penetration, vapor mixing, and post-shock heat exchange, which can further modify the thermal state of the impactor.

%%%%%%%%%%%%%%%
% Conclusion
%%%%%%%%%%%%%%%
\section{Conclusions and Implications} 
\label{sec_conclusion}

In this study, we investigated hypervelocity impacts of rocky impactors onto icy targets using three-dimensional iSALE simulations, focusing on the combined effects of porosities of both the impactor and the target. Our results demonstrate that porosity plays an important role in controlling the early-stage crater formation morphology, shock attenuation, and thermodynamic evolution of the impactor material \citep[see e.g.,][]{Kra11,Dav10}.

The simulations reveal two distinct end-member regimes \citep[see also laboratory experiments in e.g.,][]{Oka13}. When a consolidated impactor strikes a highly porous target, the impactor penetrates deeply into the target, forming a narrow penetration channel filled with high-temperature vapor generated primarily from the porous target material. In contrast, when a highly porous impactor strikes a consolidated target, the impact produces a near-surface explosive vapor expansion. When the porosities of the impactor and target are comparable, the resulting the early-stage crater formation morphology is intermediate and tends to be more hemispherical. These results demonstrate that the contrast in porosity between the colliding bodies strongly governs the impact outcome.

We further analyzed the peak pressure and peak temperature experienced by the impactor material. Focusing on the impactor material is important because micrometeoroids can act as important carriers of exogenic material that modify the surface composition. Increasing porosity in either the impactor or the target generally reduces the peak shock pressure due to energy dissipation associated with pore collapse. However, the peak temperature does not follow the same monotonic trend. When the impactor itself is porous, irreversible compaction contributes directly to internal heating, increasing the impactor temperature. Conversely, when a consolidated impactor strikes a porous target, the impactor temperature can become high not primarily through shock compression but through thermal exchange with the surrounding hot vapor produced from the target during deep penetration. As a result, although the peak pressure and temperature vary by nearly an order of magnitude depending on porosity, the impactor material is efficiently vaporized under the hypervelocity conditions considered here.

These results have important implications for hypervelocity impacts occurring in Saturn's rings. Micrometeoroids originating from non-icy sources \citep{Kem23}, such as interplanetary dust particles, are expected to be efficiently vaporized upon impact with icy ring particles, even when porosity is high. This process may contribute to the production of transient vapor and the redistribution of exogenic material within Saturnian system \citep{Hyo25}. The strong dependence of peak pressure and temperature on porosity also suggests that variations in ring-particle porosity could influence impact-induced vapor production and energy dissipation within the rings.

Future work will extend this study to investigate the dependence on impact velocity, a broader range of porosities, and impacts of icy impactors onto rocky targets. Such extensions will further clarify the role of hypervelocity impacts in shaping the physical and compositional evolution of a variety of planetary systems.

%%%%%%%%%%%%%%
% Acknowledgement
%%%%%%%%%%%%%%
\vspace{2\baselineskip}
\noindent \textbf{Acknowledgments} 
We gratefully acknowledge the developers of iSALE-3D, including Dirk Elbeshausen, Kai Wünnemann, Gareth Collins, and Tom Davison. We also thank Tom Davison, the developer of the pySALEPlot tool, which we used to analyze data in this work. iSALE-3D models were carried out on the PC cluster at the Center for Computational Astrophysics, National Astronomical Observatory of Japan. R.H. acknowledges the financial support of JSPS Grants-in-Aid (23KK0253, 22K14091, 21H04512, 21H04514, 20KK0080, 26K00756). BCJ was supported by NASA Solar Systems Workings Grant 80NSSC26K0236.

\newpage
%%%%%%%%%%%%%%
% Appendix
%%%%%%%%%%%%%%
\appendix
\renewcommand{\thefigure}{\thesection\arabic{figure}}
\setcounter{figure}{0}

%--------------------------------------%
%  SPH vs iSALE 
%--------------------------------------%
\section{Comparison between iSALE-3D and SPH-3D}
\label{sec:iSALE-vs-SPH}

Here we compare two different methods. The three-dimensional SPH hypervelocity impact simulations were performed by previous studies \citep{Hyo20,Hyo21a,Hyo25}. In these SPH simulations, the relevant EOS for water ice was a five-phase EOS \citep{Sen08}. That for non-ice materials, denoted by silicate in our study, was idealized by the quartz, SiO$_2$, M-ANEOS EOS \citep{Mel07}, as the thermodynamic data for quartz have been extensively studied and the data are sufficiently large and reliable. Here, the same composition was used for both the impactor and the target.

\begin{figure}
\begin{center}
	\includegraphics[width=1.0\textwidth]{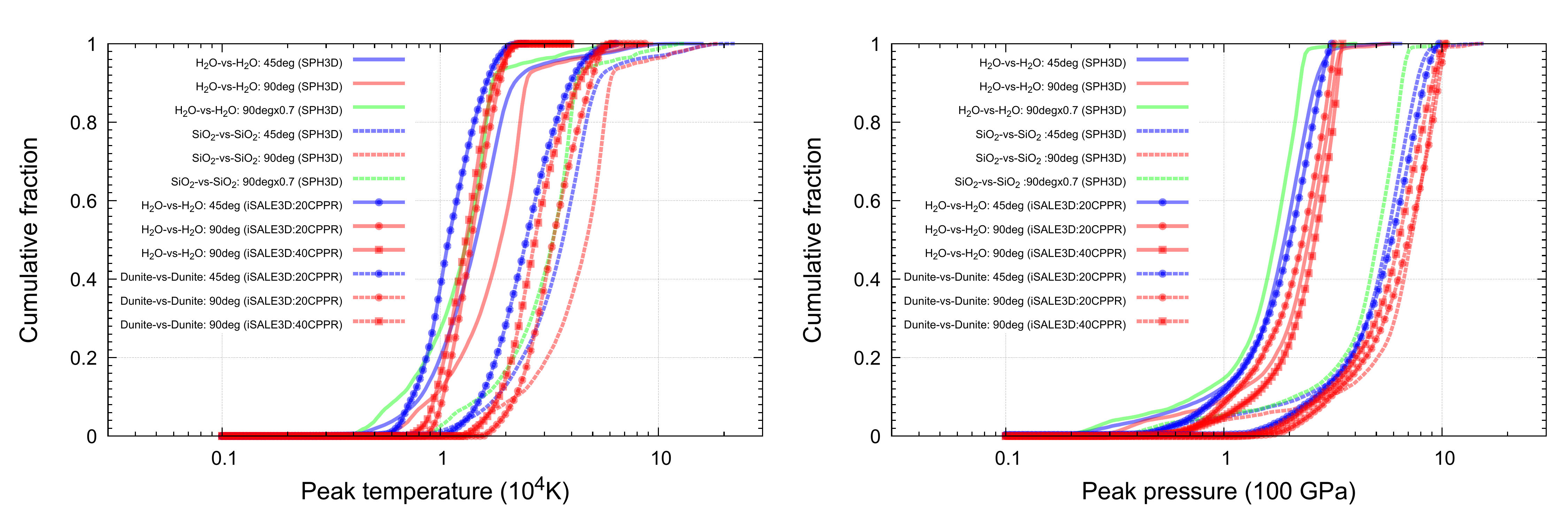}
\caption{Comparison of cumulative peak-pressure and peak-temperature distributions from SPH-3D and iSALE-3D simulations for different combinations of target and impactor compositions and for impact angles of $45^\circ$ and $90^\circ$. The results show overall consistency between the two numerical methods. We note that some discrepancies between iSALE-3D and SPH-3D, especially in the peak-temperature distributions, may originate not only from methodological differences but also from the different EOS and material models employed in the two simulations (e.g., SiO$_2$ vs. dunite). Target material is not included because the impactor material is the focus of this study. No porosity is included in these simulations.}
\label{Fig_SPH-vs-iSALE}
\end{center}
\end{figure}

Figure~\ref{Fig_SPH-vs-iSALE} shows the cumulative distributions of the peak pressure and peak temperature experienced by the impactor material during the impact process. The impact setup is basically the same as described in the Methods section (e.g., $v_{\rm imp}=30\,\mathrm{km\,s^{-1}}$), and includes different compositions (dunite and water ice) and impact angles ($45^\circ$ and $90^\circ$). We find overall consistency between the SPH-3D and iSALE-3D results, although small discrepancies remain. These discrepancies may arise not only from methodological differences, such as the presence or absence of material strength and differences in numerical schemes, but also from the different EOS models (e.g., SiO$_2$ vs dunite) employed in the two methods. We also note that, although material strength is included in our iSALE simulations, the comparison with the SPH simulations suggests that its effect does not appear to be significant in the high-velocity regime considered here, where the impactor material is strongly heated and vaporized. At lower velocities, where material strength becomes more important, the two methods may yield more divergent results \citep[e.g.,][]{Kurosawa2018,Wakita2019,Wakita2022,Man22}.

In Fig.~\ref{Fig_SPH-vs-iSALE}, the green curves show the results for $90^\circ$ impacts after scaling the horizontal axis by a factor of $0.7$. This scaling accounts for the reduction in the normal component of the impact velocity when the impact angle changes from $90^\circ$ to $45^\circ$, since $v_{\rm imp,nor} = v_{\rm imp}\sin(45^\circ) \approx 0.7v_{\rm imp}$. From an idealized analytical consideration (see Appendix Eq.~\ref{eq_Psolid}), $P_{\rm peak}$ is expected to scale as $P_{\rm peak} \propto v_{\rm imp}^2$ in the limit $S u_{\rm p} \gg C$, which applies here. This implies that changing the impact angle from $90^\circ$ to $45^\circ$ would reduce $P_{\rm peak}$ by a factor of $(0.7)^2 \approx 0.5$. However, the numerical simulations show that both $P_{\rm peak}$ and $T_{\rm peak}$ decrease by only $\sim 0.7$ rather than $\sim 0.5$, likely reflecting geometric effects and non-ideal energy partitioning not captured by the simplified analytical estimate.

%%%%%%%%%%%
% Figure S1
%%%%%%%%%%%

\newpage
%------------------------------------------------%
%  Tpeak & Ppeak: Semi-Analytical
%------------------------------------------------%
\section{Derivation of Semi-Analytical Estimation of Peak Pressure and Peak Temperature in High-Velocity Impacts}
\label{sec:derivation}

Here, we consider a one--dimensional, normal impact between two semi-infinite bodies: an \emph{impactor} (subscript $i$) and a \emph{target} (subscript $t$), composed of different materials and possibly different porosities. For each material, we define intrinsic solid matrix density, i.e., density of the pore-free material ($\rho_{s,0}$), porosity, i.e., void volume fraction ($\phi$), bulk porous initial density ($\rho_{p,0}=\rho_{s,0} (1-\phi)$), initial specific volume of the solid matrix ($V_{s,0} = 1/\rho_{s,0}$), initial specific volume of the porous aggregate ($V_{p,0} = 1/\rho_{p,0}$).

The impact velocity is denoted by $v_{\rm imp}$. After shock formation, the particle velocities behind the shock in the impactor and target are denoted by $u_{p,i}$ and $u_{p,t}$, respectively. For a planar, normal impact, kinematics imposes
\begin{equation}
  u_{p,i} + u_{p,t} = v_{\rm imp}.
  \label{eq:up_sum}
\end{equation}

We assume that peak pressure is defined as the pressure immediately behind the first compressive shock at the material interface. At the interface between the impactor and the target, the normal stress must be continuous. Therefore, the interface pressure satisfies
\begin{equation}
  P_i(u_{p,i}) = P_t(u_{p,t}),
  \label{eq:pressure_match}
\end{equation}
where $P_i$ and $P_t$ are the post-shock pressures in the impactor and target, evaluated through their respective constitutive relations.

Under the present one-dimensional, semi-infinite approximation, the solution of Eqs.~(\ref{eq:up_sum}) and (\ref{eq:pressure_match}), i.e., the impedance matching condition, yields the interface pressure immediately behind the first compressive shock, which we define as the peak pressure, $P_{\rm peak} \equiv P_i(u_{p,i}) = P_t(u_{p,t})$. This definition neglects subsequent modifications due to rarefaction waves, geometric focusing, or multi-wave shock structures.

%%%%%%%%%%%%%%%%%
\subsection{Pressure Models and Porous Compaction}
\label{subsec:pressure_models}

For solid materials (e.g., rock, ice, metals) in the shock regime, a common first-order approximation is the linear shock--particle velocity relation
\begin{equation}
  U_s = C + S u_p,
\end{equation}
where $U_s$ is the shock velocity, and $C$ and $S$ are material parameters. Using the Rankine--Hugoniot momentum condition, the post-shock solid Hugoniot pressure for the solid matrix can be written as
\begin{equation}
  P_{\rm solid} (u_p)
  = \rho_{s,0} U_s u_p
  = \rho_{s,0} (C + S u_p) u_p.
\label{eq_Psolid}
\end{equation}
Since the subsequent porous compaction model and thermodynamic integration require the pressure to be evaluated as a function of volumetric strain, it is necessary to express the solid Hugoniot pressure as an explicit function of the specific volume $V_s$. By combining Equation (\ref{eq_Psolid}) with the Rankine--Hugoniot mass conservation relation ($1 - V_s/V_{s,0} = u_p/U_s$), we obtain
\begin{equation}
  P_{\rm solid}(V_s) = \frac{C^2(V_{s,0} - V_s)}{[V_{s,0} - S(V_{s,0} - V_s)]^2}.
\label{eq_Psolid_Vs}
\end{equation}

To extend our formulation to porous materials, we adopt the concept of the distension, $\alpha$, defined as the ratio of the specific volume of the porous aggregate ($V_p$) to that of the solid matrix ($V_s$):
\begin{equation}
\label{eq:alpha}
  \alpha \equiv \frac{V_p}{V_s}.
\end{equation}
The initial value, prior to shock-induced compaction, is
\begin{equation}
\label{eq:alpha_0}
  \alpha_0 = \frac{V_{p,0}}{V_{s,0} } = \frac{1}{1-\phi}.
\end{equation}

Using $\alpha$, the bulk porous pressure is related to the solid matrix pressure by the Carroll-Holt relation \citep{Car72}:
\begin{equation}
  P_{\rm porous}(V_s) = \frac{1}{\alpha} P_{\rm solid}(V_s), \quad \text{where} \quad V_s = \frac{V_p}{\alpha}.
  \label{eq:porous_pressure}
\end{equation}
Here, $P_{\rm solid}(V_s)$ is the pressure of the non-porous material at the specific volume $V_s$, evaluated along the solid Hugoniot (Eq.~\ref{eq_Psolid_Vs}). More generally, $P_{\rm solid}(V_s)$ may be obtained by direct evaluation from a tabulated or analytic EOS (e.g., Mie--Gr\"uneisen, ANEOS) at the state $(V_s, E_s)$.

%%%%%%%%%%%%%%%%%
\subsection{$\varepsilon$--$\alpha$ Model for Porous Compaction}

Here, we adopt the $\varepsilon$--$\alpha$ compaction model, which describes the reduction of porosity as a function of volumetric strain ($\epsilon \equiv \ln \left( V_{p} / V_{p,0} \right) $) rather than pressure. This approach is also adopted in the iSALE shock physics code \citep{Wunnemann2006}.

The distension $\alpha$ is determined by the bulk compression of the material. We utilize the simplified power-law form for the crush curve:
\begin{equation}
  \alpha(V_p) = \max\left(1, \, \alpha_0 \left( \frac{V_p}{V_{p,0}} \right)^\kappa \right),
  \label{eq:epsilon_alpha}
\end{equation}
where $\kappa$ is the compaction parameter (typically $\kappa \approx 0.98$ for geological materials) describing the efficiency of pore collapse.
According to Eq.~(\ref{eq:epsilon_alpha}), as the bulk volume $V_p$ decreases from $V_{p,0}$ during shock compression, $\alpha$ decreases from $\alpha_0$ until it reaches unity (full compaction). Unlike the $p$--$\alpha$ model, where $\alpha$ depends implicitly on pressure, here $\alpha$ is an explicit function of the current volume $V_p$. This simplifies the numerical solution by allowing $\alpha$ to be determined directly from the geometric state.

%%%%%%%%%%%%%%%%%
\subsection{Energy Partition in Porous Materials}
For porous materials, it is useful to decompose the total mechanical work into two physically distinct contributions. We define the total specific energy imparted to the porous aggregate as
\begin{equation}
\label{eq:Etot}
  E_{\text{total}}
  = \frac{1}{2} P \left( V_{p,0} - V_p \right),
\end{equation}
where $P$ denotes the shock pressure of the final compressed porous state, obtained from the intersection between the Rayleigh line and the porous Hugoniot in Step 2 below. It is not the pressure immediately before porosity changes, but an effective Hugoniot pressure used in the Rankine--Hugoniot energy relation between the initial porous state $V_{p,0}$ and the final compressed state $V_p$.

Alternatively, the energy associated with compressing only the solid matrix along its reference Hugoniot is
\begin{equation}
  E_{\text{solid}}
  = \frac{1}{2} P_{\text{solid}}(V_s) \left( V_{s,0}  - V_s \right).
\end{equation}
The irreversible contribution associated with pore collapse (i.e., waste energy) is obtained by the energy balance:
\begin{equation}
\label{eq:waste_energy}
  E_{\text{waste}}
  = E_{\text{total}} - E_{\text{solid}}.
\end{equation}
%

%%%%%%%%%%%%%%%%%
\subsection{Temperature Estimation and Solution Procedure}

The peak post-shock temperature is approximated as the sum of the isentropic/Hugoniot temperature of the solid matrix and the temperature rise due to waste energy using a material-dependent constant specific heat $C_v$:
\begin{equation}
  T_{\text{peak}} = T_{\text{solid}}(V_s) + \frac{E_{\text{waste}}}{C_v},
\end{equation}
where $T_{\text{solid}}(V_s)$ is obtained by integrating the thermodynamic relation along the reference solid Hugoniot:
\begin{equation}
	\frac{dT}{dV_s} = -\frac{\gamma}{V_s} T + \frac{1}{2 C_v} \left[ P_{\text{solid}}(V_s) + (V_{s,0}  - V_s) \frac{dP_{\text{solid}}}{dV_s} \right].
\end{equation}

Therefore, for a given impact velocity $v_{\rm imp}$ and material properties $(\rho_{s,0}, \phi, C, S, \gamma, C_v, \kappa)$, the peak state is obtained as follows:

\begin{enumerate}[Step \arabic*:]
\item Impedance Matching: Solve Eq.~(\ref{eq:up_sum}) and (\ref{eq:pressure_match}) to find the common interface pressure $P_{\rm peak}$ and particle velocities $u_{\rm p}$. This requires evaluating the porous Hugoniot pressure $P(u_p)$ for both impactor and target at each trial $u_{\rm p}$ generated by the root-finding algorithm.

\item Porous Hugoniot Evaluation: During the impedance-matching iteration in Step 1, for any trial particle velocity $u_p$, we determine the corresponding compressed porous volume $V_p$ from the Rankine--Hugoniot mass and momentum conditions. The momentum condition gives $P = \rho_{p,0}U_su_p$, while mass conservation gives $V_p/V_{p,0}=1-u_p/U_s$, or $U_s=V_{p,0}u_p/(V_{p,0}-V_p)$. Eliminating $U_s$ yields the Rayleigh-line pressure
\begin{equation}
P_{\rm Rayleigh}(V_p)
= \rho_{p,0} u_p^2
\frac{V_{p,0}}{V_{p,0}-V_p}.
\end{equation}
The physically allowed shocked state is obtained where this Rayleigh line intersects the porous Hugoniot predicted by the compaction model:
\begin{equation}
P_{\rm Rayleigh}(V_p) = P_{\rm model}(V_p),
\end{equation}
where $P_{\rm model}(V_p) \equiv P_{\rm solid}(V_p/\alpha)/\alpha$ and $\alpha=\alpha(V_p)$ is evaluated from Eq.~(\ref{eq:epsilon_alpha}).

\end{enumerate}

\noindent In the $\varepsilon$--$\alpha$ model, this evaluation is robust because for any trial volume $V_p$:
  \begin{enumerate}
  \renewcommand{\labelenumi}{\alph{enumi})} 
    \item Compute distension $\alpha(V_p)$ directly using Eq.~(\ref{eq:epsilon_alpha}).
    \item Compute matrix volume $V_s = V_p / \alpha$.
    \item Evaluate solid pressure $P_{\rm solid}(V_s)$ and obtain $P_{\rm model} = P_{\rm solid} / \alpha$.
    \item Iterate on $V_p$ using a root-finding algorithm (e.g., Brent's method) until the Rayleigh line pressure matches the model pressure.
\end{enumerate}

\begin{enumerate}[Step \arabic*:]
\setcounter{enumi}{2}
\item Energy and Temperature: Once the equilibrium state $(P, V_p)$ is found, compute $E_{\text{total}}$, $E_{\text{solid}}$, $E_{\text{waste}}$, and finally $T_{\text{peak}}$.
\end{enumerate}

Lastly, we briefly explain how the porosity dependence appears explicitly in the analytical estimate. To clarify the physical origin of the enhanced temperature rise in high-porosity cases, we consider a simplified limiting case in which pore collapse is effectively complete. In this limit, the compressed porous volume approaches the matrix volume, $V_p \approx V_s$, and the porous-state pressure $P$, defined in Eq.~\ref{eq:Etot}, is equal to the pressure at the intersection between the Rayleigh line and the porous Hugoniot, i.e., $P=P_{\rm Rayleigh}(V_p)=P_{\rm model}(V_p)$.

Under these assumptions, the irreversible waste energy defined in Eq.~(\ref{eq:waste_energy}) can be approximated as
\begin{align}
    E_{\rm waste}
    &= E_{\rm total} - E_{\rm solid} \nonumber \\
    &\approx \frac{1}{2} P (V_{p,0} - V_s)
    - \frac{1}{2} P (V_{s,0} - V_s) \nonumber \\
    &= \frac{1}{2} P (V_{p,0} - V_{s,0}) .
\end{align}
Using $V_{p,0}=V_{s,0}/(1-\phi)$ and $V_{s,0}=1/\rho_{s,0}$, this becomes
\begin{align}
    E_{\rm waste}
    &\approx
    \frac{1}{2} P
    \left[
    \frac{V_{s,0}}{1-\phi} - V_{s,0}
    \right] \nonumber \\
    &=
    \frac{P}{2\rho_{s,0}}
    \left(
    \frac{\phi}{1-\phi}
    \right) .
\end{align}
Therefore, in this fully compacted limiting case, the temperature rise associated with irreversible pore-collapse work, $\Delta T \approx E_{\rm waste}/C_v$, contains an explicit nonlinear dependence on porosity through the factor $\phi/(1-\phi)$. In the full semi-analytical model, this tendency is moderated by the simultaneous porosity dependence of $P$, because increasing porosity also lowers the shock pressure through the Rayleigh-line and impedance-matching conditions.

%--------------------------------------%
%  Hugoniot
%--------------------------------------%
\setcounter{figure}{0}
\section{Porosity Dependence of Hugoniot Curves and Vaporization Thresholds}
\label{appendix_Hugoniot}

%%%%%%%%%%%
% Figure A2
%%%%%%%%%%%
\begin{figure}
\begin{center}
	\includegraphics[width=0.7\textwidth]{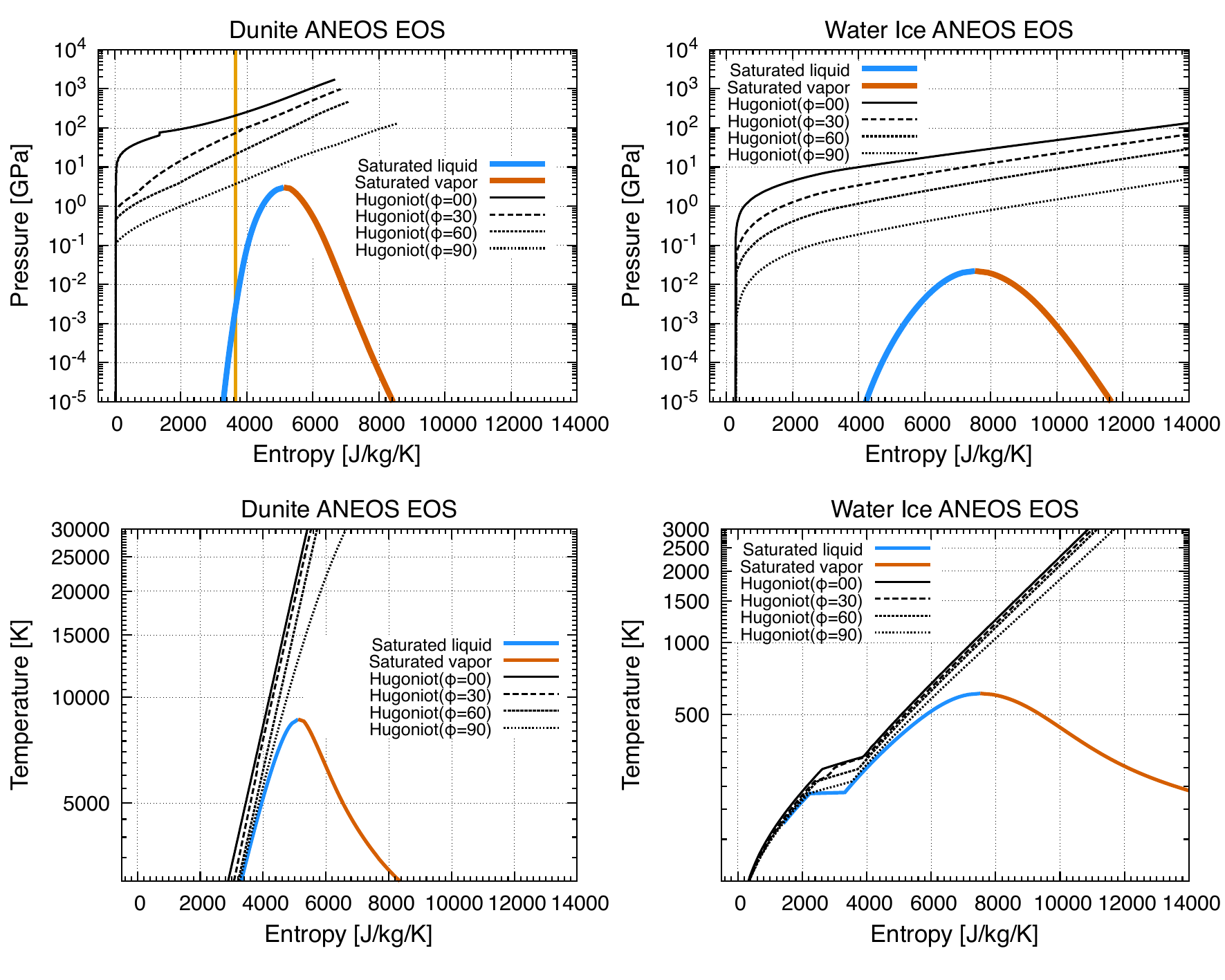}
\caption{Pressure--entropy and temperature--entropy diagrams for the adopted ANEOS equations of state for dunite and water ice. Black curves show Hugoniot curves calculated for different initial porosities, $\phi=0$, 30, 60, and 90\%. Blue and red curves indicate the saturated liquid and saturated vapor curves, respectively. In the upper-left panel, the orange vertical line marks the critical entropy for incipient vaporization of dunite, $S=3663$\,J\,kg$^{-1}$\,K$^{-1}$ \citep{Pie97}. The intersections of this entropy criterion with the dunite Hugoniot curves indicate the corresponding critical shock pressures required for incipient vaporization. For highly porous dunite, this pressure is substantially lower than for non-porous dunite; for $\phi=90\%$ and $\phi=0\%$, the corresponding pressures are approximately $3$\,GPa and $200$\,GPa, respectively. Note that the dunite Hugoniot curves in the upper-left panel are shown only over the limited entropy range for which stable EOS outputs were numerically obtained. The water-ice panels show that the relevant saturation and Hugoniot curves lie at substantially lower pressures than those for dunite, indicating that water ice enters the vaporization regime at much lower shock pressures. The bottom panels show the corresponding temperature--entropy relations and are provided as reference information for the adopted EOSs.}
\label{Fig_S-P-T}
\end{center}
\end{figure}

Here, we examine the porosity dependence of the Hugoniot curves used to evaluate the vaporization state of the impactor material. Figure~\ref{Fig_S-P-T} shows pressure--entropy and temperature--entropy diagrams for the adopted ANEOS equations of state for dunite and water ice. The black curves show the principal Hugoniot curves calculated for different initial porosities, while the blue and red curves indicate the saturated liquid and saturated vapor curves, respectively. In the upper-left panel, the orange vertical line marks the critical entropy for incipient vaporization of dunite, $S=3663$\,J\,kg$^{-1}$\,K$^{-1}$, reported by \citet{Pie97}.

The Hugoniot curves depend strongly on the initial porosity. For a given entropy, higher initial porosity corresponds to a substantially lower shock pressure \citep[see also][]{Kra11,Ste08}. This behavior arises because pore collapse converts a larger fraction of the shock work into irreversible internal heating, thereby increasing the entropy at lower pressure than in the non-porous case. Thus, the intersections between the orange entropy criterion and the dunite Hugoniot curves provide the shock pressures required for incipient vaporization for each initial porosity.

In this study, we use the critical entropy of $3663$\,J\,kg$^{-1}$\,K$^{-1}$ as the incipient-vaporization criterion for dunite \citep{Pie97}. For the 90\% porous dunite Hugoniot, the pressure corresponding to this entropy is approximately $3$\,GPa, whereas the corresponding pressure for non-porous dunite is nearly two orders of magnitude larger (i.e., $\sim 200$\,GPa). We therefore adopt $\sim 3$\,GPa as the nominal incipient-vaporization pressure threshold for 90\% porous dunite in this study. This threshold should be regarded as a conservative pressure-based proxy, because it is derived from the intersection of the entropy criterion with the Hugoniot curve and does not include additional heating mechanisms such as shear/plastic deformation heating \citep[e.g.,][]{Kurosawa2018,Wakita2019} or post-shock thermal exchange with hot target-derived material, as observed in our simulations. Therefore, the use of the dunite-based incipient-vaporization threshold provides a conservative basis for assessing vaporization of the rocky impactor, while the vaporization of water ice should be interpreted qualitatively from the EOS behavior shown in Fig.~\ref{Fig_S-P-T}.

It should also be noted that the Hugoniot-based criterion discussed here accounts only for vaporization directly associated with impact-shock compression. In the simulations presented in this study, the impactor material can also be heated during post-shock evolution through thermal exchange with hot target-derived material, particularly when a consolidated impactor penetrates into a highly porous icy target (see Sec.~\ref{sec:Ppeak_Tpeak}). This post-shock heating mechanism is distinct from both direct shock heating and shear/plastic deformation heating, and is not included in the pressure--entropy estimate shown in Fig.~\ref{Fig_S-P-T}. Therefore, the pressure threshold inferred from the dunite Hugoniot should be regarded as a conservative criterion: impactor vaporization may occur even when the peak pressure alone is lower than the nominal Hugoniot-based threshold.

For water ice, we do not introduce a separate reference entropy for incipient vaporization in this study, because our main focus is the thermodynamic fate of the rocky impactor. Nevertheless, the water-ice panels in Fig.~\ref{Fig_S-P-T} show that the saturation curves and Hugoniot curves occur at substantially lower pressures than those for dunite \citep[see also][]{Kra11,Ste08}. This qualitative comparison indicates that water ice can enter the vaporization regime at much lower shock pressures than dunite.

The bottom panels of Fig.~\ref{Fig_S-P-T} show the corresponding temperature--entropy relations along the same Hugoniot curves. These temperature--entropy diagrams are not used directly to define the vaporization criterion in this study, but are provided as reference information for the adopted EOSs. They illustrate how the post-shock temperature varies with entropy and initial porosity for dunite and water ice. In particular, higher-porosity Hugoniot curves generally reach higher entropies at comparable thermodynamic conditions, reflecting additional irreversible heating associated with pore collapse. Thus, the bottom panels provide complementary thermodynamic context for the pressure--entropy diagrams used above.

\newpage
\printcredits

\noindent \textbf{Declaration of generative AI and AI-assisted technologies in the manuscript preparation process}
During the preparation of this work the author(s) used ChatGPT in order to correct English. After using this tool/service, the author(s) reviewed and edited the content as needed and take(s) full responsibility for the content of the published article.

%%%%%%%%%%%
%  bibliography
%%%%%%%%%%%
\newpage
\bibliographystyle{cas-model2-names}
\bibliography{ref}

\begin{thebibliography}{54}
\expandafter\ifx\csname natexlab\endcsname\relax\def\natexlab#1{#1}\fi
\providecommand{\url}[1]{\texttt{#1}}
\providecommand{\href}[2]{#2}
\providecommand{\path}[1]{#1}
\providecommand{\DOIprefix}{doi:}
\providecommand{\ArXivprefix}{arXiv:}
\providecommand{\URLprefix}{URL: }
\providecommand{\Pubmedprefix}{pmid:}
\providecommand{\doi}[1]{\href{http://dx.doi.org/#1}{\path{#1}}}
\providecommand{\Pubmed}[1]{\href{pmid:#1}{\path{#1}}}
\providecommand{\bibinfo}[2]{#2}
\ifx\xfnm\relax \def\xfnm[#1]{\unskip,\space#1}\fi
%Type = Article
\bibitem[{{Altobelli} et~al.(2019){Altobelli}, {Fiege}, {Carry}, {Soja},
  {Guglielmino}, {Trieloff}, {Orlando} and {Srama}}]{Alt19}
\bibinfo{author}{{Altobelli}, N.}, \bibinfo{author}{{Fiege}, K.},
  \bibinfo{author}{{Carry}, B.}, \bibinfo{author}{{Soja}, R.},
  \bibinfo{author}{{Guglielmino}, M.}, \bibinfo{author}{{Trieloff}, M.},
  \bibinfo{author}{{Orlando}, T.M.}, \bibinfo{author}{{Srama}, R.},
  \bibinfo{year}{2019}.
\newblock \bibinfo{title}{{Space Weathering Induced Via Microparticle Impacts:
  1. Modeling of Impact Velocities and Flux of Micrometeoroids From Cometary,
  Asteroidal, and Interstellar Origin in the Main Asteroid Belt and the
  Near-Earth Environment}}.
\newblock \bibinfo{journal}{Journal of Geophysical Research (Planets)}
  \bibinfo{volume}{124}, \bibinfo{pages}{1044--1083}.
\newblock \DOIprefix\doi{10.1029/2018JE005563}.
%Type = Article
\bibitem[{{Babadzhanov} and {Kokhirova}(2009)}]{Bab09}
\bibinfo{author}{{Babadzhanov}, P.B.}, \bibinfo{author}{{Kokhirova}, G.I.},
  \bibinfo{year}{2009}.
\newblock \bibinfo{title}{{Densities and porosities of meteoroids}}.
\newblock \bibinfo{journal}{\aap} \bibinfo{volume}{495},
  \bibinfo{pages}{353--358}.
\newblock \DOIprefix\doi{10.1051/0004-6361:200810460}.
%Type = Article
\bibitem[{{Benz} et~al.(1989){Benz}, {Cameron} and {Melosh}}]{Benz1989}
\bibinfo{author}{{Benz}, W.}, \bibinfo{author}{{Cameron}, A.G.W.},
  \bibinfo{author}{{Melosh}, H.J.}, \bibinfo{year}{1989}.
\newblock \bibinfo{title}{{The origin of the Moon and the single-impact
  hypothesis III}}.
\newblock \bibinfo{journal}{Icarus} \bibinfo{volume}{81},
  \bibinfo{pages}{113--131}.
\newblock \DOIprefix\doi{10.1016/0019-1035(89)90129-2}.
%Type = Article
\bibitem[{{Blum}(2006)}]{Blu06}
\bibinfo{author}{{Blum}, J.}, \bibinfo{year}{2006}.
\newblock \bibinfo{title}{{Dust agglomeration}}.
\newblock \bibinfo{journal}{Advances in Physics} \bibinfo{volume}{55},
  \bibinfo{pages}{881--947}.
\newblock \DOIprefix\doi{10.1080/00018730601095039}.
%Type = Article
\bibitem[{{Blum} et~al.(2022){Blum}, {Bischoff} and {Gundlach}}]{Blu22}
\bibinfo{author}{{Blum}, J.}, \bibinfo{author}{{Bischoff}, D.},
  \bibinfo{author}{{Gundlach}, B.}, \bibinfo{year}{2022}.
\newblock \bibinfo{title}{{Formation of Comets}}.
\newblock \bibinfo{journal}{Universe} \bibinfo{volume}{8},
  \bibinfo{pages}{381}.
\newblock \DOIprefix\doi{10.3390/universe8070381},
  \href{http://arxiv.org/abs/2207.12731}{\tt arXiv:2207.12731}.
%Type = Article
\bibitem[{{Blum} and {Wurm}(2008)}]{Blu08}
\bibinfo{author}{{Blum}, J.}, \bibinfo{author}{{Wurm}, G.},
  \bibinfo{year}{2008}.
\newblock \bibinfo{title}{{The growth mechanisms of macroscopic bodies in
  protoplanetary disks.}}
\newblock \bibinfo{journal}{Annu. Rev. Astron. Astrophys.}
  \bibinfo{volume}{46}, \bibinfo{pages}{21--56}.
\newblock \DOIprefix\doi{10.1146/annurev.astro.46.060407.145152}.
%Type = Article
\bibitem[{{Bray} et~al.(2014){Bray}, {Collins}, {Morgan}, {Melosh} and
  {Schenk}}]{Bray2014}
\bibinfo{author}{{Bray}, V.J.}, \bibinfo{author}{{Collins}, G.S.},
  \bibinfo{author}{{Morgan}, J.V.}, \bibinfo{author}{{Melosh}, H.J.},
  \bibinfo{author}{{Schenk}, P.M.}, \bibinfo{year}{2014}.
\newblock \bibinfo{title}{{Hydrocode simulation of Ganymede and Europa
  cratering trends - How thick is Europa{\textquoteright}s crust?}}
\newblock \bibinfo{journal}{Icarus} \bibinfo{volume}{231},
  \bibinfo{pages}{394--406}.
\newblock \DOIprefix\doi{10.1016/j.icarus.2013.12.009}.
%Type = Article
\bibitem[{{Brownlee} et~al.(2006){Brownlee}, {Tsou}, {Al{\'e}on}, {Alexander},
  {Araki}, {Bajt}, {Baratta}, {Bastien}, {Bland}, {Bleuet}, {Borg}, {Bradley},
  {Brearley}, {Brenker}, {Brennan}, {Bridges}, {Browning}, {Brucato},
  {Bullock}, {Burchell}, {Busemann}, {Butterworth}, {Chaussidon}, {Cheuvront},
  {Chi}, {Cintala}, {Clark}, {Clemett}, {Cody}, {Colangeli}, {Cooper},
  {Cordier}, {Daghlian}, {Dai}, {D'Hendecourt}, {Djouadi}, {Dominguez},
  {Duxbury}, {Dworkin}, {Ebel}, {Economou}, {Fakra}, {Fairey}, {Fallon},
  {Ferrini}, {Ferroir}, {Fleckenstein}, {Floss}, {Flynn}, {Franchi}, {Fries},
  {Gainsforth}, {Gallien}, {Genge}, {Gilles}, {Gillet}, {Gilmour}, {Glavin},
  {Gounelle}, {Grady}, {Graham}, {Grant}, {Green}, {Grossemy}, {Grossman},
  {Grossman}, {Guan}, {Hagiya}, {Harvey}, {Heck}, {Herzog}, {Hoppe},
  {H{\"o}rz}, {Huth}, {Hutcheon}, {Ignatyev}, {Ishii}, {Ito}, {Jacob},
  {Jacobsen}, {Jacobsen}, {Jones}, {Joswiak}, {Jurewicz}, {Kearsley}, {Keller},
  {Khodja}, {Kilcoyne}, {Kissel}, {Krot}, {Langenhorst}, {Lanzirotti}, {Le},
  {Leshin}, {Leitner}, {Lemelle}, {Leroux}, {Liu}, {Luening}, {Lyon},
  {MacPherson}, {Marcus}, {Marhas}, {Marty}, {Matrajt}, {McKeegan}, {Meibom},
  {Mennella}, {Messenger}, {Messenger}, {Mikouchi}, {Mostefaoui}, {Nakamura},
  {Nakano}, {Newville}, {Nittler}, {Ohnishi}, {Ohsumi}, {Okudaira},
  {Papanastassiou}, {Palma}, {Palumbo}, {Pepin}, {Perkins}, {Perronnet},
  {Pianetta}, {Rao}, {Rietmeijer}, {Robert}, {Rost}, {Rotundi}, {Ryan},
  {Sandford}, {Schwandt}, {See}, {Schlutter}, {Sheffield-Parker},
  {Simionovici}, {Simon}, {Sitnitsky}, {Snead}, {Spencer}, {Stadermann},
  {Steele}, {Stephan}, {Stroud}, {Susini}, {Sutton}, {Suzuki}, {Taheri},
  {Taylor}, {Teslich}, {Tomeoka}, {Tomioka}, {Toppani}, {Trigo-Rodr{\'\i}guez},
  {Troadec}, {Tsuchiyama}, {Tuzzolino}, {Tyliszczak}, {Uesugi}, {Velbel},
  {Vellenga}, {Vicenzi}, {Vincze}, {Warren}, {Weber}, {Weisberg}, {Westphal},
  {Wirick}, {Wooden}, {Wopenka}, {Wozniakiewicz}, {Wright}, {Yabuta}, {Yano},
  {Young}, {Zare}, {Zega}, {Ziegler}, {Zimmerman}, {Zinner} and
  {Zolensky}}]{Bro06}
\bibinfo{author}{{Brownlee}, D.}, \bibinfo{author}{{Tsou}, P.},
  \bibinfo{author}{{Al{\'e}on}, J.}, \bibinfo{author}{{Alexander}, C.M.O.D.},
  \bibinfo{author}{{Araki}, T.}, \bibinfo{author}{{Bajt}, S.},
  \bibinfo{author}{{Baratta}, G.A.}, \bibinfo{author}{{Bastien}, R.},
  \bibinfo{author}{{Bland}, P.}, \bibinfo{author}{{Bleuet}, P.},
  \bibinfo{author}{{Borg}, J.}, \bibinfo{author}{{Bradley}, J.P.},
  \bibinfo{author}{{Brearley}, A.}, \bibinfo{author}{{Brenker}, F.},
  \bibinfo{author}{{Brennan}, S.}, \bibinfo{author}{{Bridges}, J.C.},
  \bibinfo{author}{{Browning}, N.D.}, \bibinfo{author}{{Brucato}, J.R.},
  \bibinfo{author}{{Bullock}, E.}, \bibinfo{author}{{Burchell}, M.J.},
  \bibinfo{author}{{Busemann}, H.}, \bibinfo{author}{{Butterworth}, A.},
  \bibinfo{author}{{Chaussidon}, M.}, \bibinfo{author}{{Cheuvront}, A.},
  \bibinfo{author}{{Chi}, M.}, \bibinfo{author}{{Cintala}, M.J.},
  \bibinfo{author}{{Clark}, B.C.}, \bibinfo{author}{{Clemett}, S.J.},
  \bibinfo{author}{{Cody}, G.}, \bibinfo{author}{{Colangeli}, L.},
  \bibinfo{author}{{Cooper}, G.}, \bibinfo{author}{{Cordier}, P.},
  \bibinfo{author}{{Daghlian}, C.}, \bibinfo{author}{{Dai}, Z.},
  \bibinfo{author}{{D'Hendecourt}, L.}, \bibinfo{author}{{Djouadi}, Z.},
  \bibinfo{author}{{Dominguez}, G.}, \bibinfo{author}{{Duxbury}, T.},
  \bibinfo{author}{{Dworkin}, J.P.}, \bibinfo{author}{{Ebel}, D.S.},
  \bibinfo{author}{{Economou}, T.E.}, \bibinfo{author}{{Fakra}, S.},
  \bibinfo{author}{{Fairey}, S.A.J.}, \bibinfo{author}{{Fallon}, S.},
  \bibinfo{author}{{Ferrini}, G.}, \bibinfo{author}{{Ferroir}, T.},
  \bibinfo{author}{{Fleckenstein}, H.}, \bibinfo{author}{{Floss}, C.},
  \bibinfo{author}{{Flynn}, G.}, \bibinfo{author}{{Franchi}, I.A.},
  \bibinfo{author}{{Fries}, M.}, \bibinfo{author}{{Gainsforth}, Z.},
  \bibinfo{author}{{Gallien}, J.P.}, \bibinfo{author}{{Genge}, M.},
  \bibinfo{author}{{Gilles}, M.K.}, \bibinfo{author}{{Gillet}, P.},
  \bibinfo{author}{{Gilmour}, J.}, \bibinfo{author}{{Glavin}, D.P.},
  \bibinfo{author}{{Gounelle}, M.}, \bibinfo{author}{{Grady}, M.M.},
  \bibinfo{author}{{Graham}, G.A.}, \bibinfo{author}{{Grant}, P.G.},
  \bibinfo{author}{{Green}, S.F.}, \bibinfo{author}{{Grossemy}, F.},
  \bibinfo{author}{{Grossman}, L.}, \bibinfo{author}{{Grossman}, J.N.},
  \bibinfo{author}{{Guan}, Y.}, \bibinfo{author}{{Hagiya}, K.},
  \bibinfo{author}{{Harvey}, R.}, \bibinfo{author}{{Heck}, P.},
  \bibinfo{author}{{Herzog}, G.F.}, \bibinfo{author}{{Hoppe}, P.},
  \bibinfo{author}{{H{\"o}rz}, F.}, \bibinfo{author}{{Huth}, J.},
  \bibinfo{author}{{Hutcheon}, I.D.}, \bibinfo{author}{{Ignatyev}, K.},
  \bibinfo{author}{{Ishii}, H.}, \bibinfo{author}{{Ito}, M.},
  \bibinfo{author}{{Jacob}, D.}, \bibinfo{author}{{Jacobsen}, C.},
  \bibinfo{author}{{Jacobsen}, S.}, \bibinfo{author}{{Jones}, S.},
  \bibinfo{author}{{Joswiak}, D.}, \bibinfo{author}{{Jurewicz}, A.},
  \bibinfo{author}{{Kearsley}, A.T.}, \bibinfo{author}{{Keller}, L.P.},
  \bibinfo{author}{{Khodja}, H.}, \bibinfo{author}{{Kilcoyne}, A.L.D.},
  \bibinfo{author}{{Kissel}, J.}, \bibinfo{author}{{Krot}, A.},
  \bibinfo{author}{{Langenhorst}, F.}, \bibinfo{author}{{Lanzirotti}, A.},
  \bibinfo{author}{{Le}, L.}, \bibinfo{author}{{Leshin}, L.A.},
  \bibinfo{author}{{Leitner}, J.}, \bibinfo{author}{{Lemelle}, L.},
  \bibinfo{author}{{Leroux}, H.}, \bibinfo{author}{{Liu}, M.C.},
  \bibinfo{author}{{Luening}, K.}, \bibinfo{author}{{Lyon}, I.},
  \bibinfo{author}{{MacPherson}, G.}, \bibinfo{author}{{Marcus}, M.A.},
  \bibinfo{author}{{Marhas}, K.}, \bibinfo{author}{{Marty}, B.},
  \bibinfo{author}{{Matrajt}, G.}, \bibinfo{author}{{McKeegan}, K.},
  \bibinfo{author}{{Meibom}, A.}, \bibinfo{author}{{Mennella}, V.},
  \bibinfo{author}{{Messenger}, K.}, \bibinfo{author}{{Messenger}, S.},
  \bibinfo{author}{{Mikouchi}, T.}, \bibinfo{author}{{Mostefaoui}, S.},
  \bibinfo{author}{{Nakamura}, T.}, \bibinfo{author}{{Nakano}, T.},
  \bibinfo{author}{{Newville}, M.}, \bibinfo{author}{{Nittler}, L.R.},
  \bibinfo{author}{{Ohnishi}, I.}, \bibinfo{author}{{Ohsumi}, K.},
  \bibinfo{author}{{Okudaira}, K.}, \bibinfo{author}{{Papanastassiou}, D.A.},
  \bibinfo{author}{{Palma}, R.}, \bibinfo{author}{{Palumbo}, M.E.},
  \bibinfo{author}{{Pepin}, R.O.}, \bibinfo{author}{{Perkins}, D.},
  \bibinfo{author}{{Perronnet}, M.}, \bibinfo{author}{{Pianetta}, P.},
  \bibinfo{author}{{Rao}, W.}, \bibinfo{author}{{Rietmeijer}, F.J.M.},
  \bibinfo{author}{{Robert}, F.}, \bibinfo{author}{{Rost}, D.},
  \bibinfo{author}{{Rotundi}, A.}, \bibinfo{author}{{Ryan}, R.},
  \bibinfo{author}{{Sandford}, S.A.}, \bibinfo{author}{{Schwandt}, C.S.},
  \bibinfo{author}{{See}, T.H.}, \bibinfo{author}{{Schlutter}, D.},
  \bibinfo{author}{{Sheffield-Parker}, J.}, \bibinfo{author}{{Simionovici},
  A.}, \bibinfo{author}{{Simon}, S.}, \bibinfo{author}{{Sitnitsky}, I.},
  \bibinfo{author}{{Snead}, C.J.}, \bibinfo{author}{{Spencer}, M.K.},
  \bibinfo{author}{{Stadermann}, F.J.}, \bibinfo{author}{{Steele}, A.},
  \bibinfo{author}{{Stephan}, T.}, \bibinfo{author}{{Stroud}, R.},
  \bibinfo{author}{{Susini}, J.}, \bibinfo{author}{{Sutton}, S.R.},
  \bibinfo{author}{{Suzuki}, Y.}, \bibinfo{author}{{Taheri}, M.},
  \bibinfo{author}{{Taylor}, S.}, \bibinfo{author}{{Teslich}, N.},
  \bibinfo{author}{{Tomeoka}, K.}, \bibinfo{author}{{Tomioka}, N.},
  \bibinfo{author}{{Toppani}, A.}, \bibinfo{author}{{Trigo-Rodr{\'\i}guez},
  J.M.}, \bibinfo{author}{{Troadec}, D.}, \bibinfo{author}{{Tsuchiyama}, A.},
  \bibinfo{author}{{Tuzzolino}, A.J.}, \bibinfo{author}{{Tyliszczak}, T.},
  \bibinfo{author}{{Uesugi}, K.}, \bibinfo{author}{{Velbel}, M.},
  \bibinfo{author}{{Vellenga}, J.}, \bibinfo{author}{{Vicenzi}, E.},
  \bibinfo{author}{{Vincze}, L.}, \bibinfo{author}{{Warren}, J.},
  \bibinfo{author}{{Weber}, I.}, \bibinfo{author}{{Weisberg}, M.},
  \bibinfo{author}{{Westphal}, A.J.}, \bibinfo{author}{{Wirick}, S.},
  \bibinfo{author}{{Wooden}, D.}, \bibinfo{author}{{Wopenka}, B.},
  \bibinfo{author}{{Wozniakiewicz}, P.}, \bibinfo{author}{{Wright}, I.},
  \bibinfo{author}{{Yabuta}, H.}, \bibinfo{author}{{Yano}, H.},
  \bibinfo{author}{{Young}, E.D.}, \bibinfo{author}{{Zare}, R.N.},
  \bibinfo{author}{{Zega}, T.}, \bibinfo{author}{{Ziegler}, K.},
  \bibinfo{author}{{Zimmerman}, L.}, \bibinfo{author}{{Zinner}, E.},
  \bibinfo{author}{{Zolensky}, M.}, \bibinfo{year}{2006}.
\newblock \bibinfo{title}{{Comet 81P/Wild 2 Under a Microscope}}.
\newblock \bibinfo{journal}{Science} \bibinfo{volume}{314},
  \bibinfo{pages}{1711}.
\newblock \DOIprefix\doi{10.1126/science.1135840}.
%Type = Article
\bibitem[{{Carroll} and {Holt}(1972)}]{Car72}
\bibinfo{author}{{Carroll}, M.M.}, \bibinfo{author}{{Holt}, A.C.},
  \bibinfo{year}{1972}.
\newblock \bibinfo{title}{{Static and Dynamic Pore-Collapse Relations for
  Ductile Porous Materials}}.
\newblock \bibinfo{journal}{Journal of Applied Physics} \bibinfo{volume}{43},
  \bibinfo{pages}{1626--1636}.
\newblock \DOIprefix\doi{10.1063/1.1661372}.
%Type = Article
\bibitem[{Collins et~al.(2011)Collins, Melosh and Wünnemann}]{collins2011}
\bibinfo{author}{Collins, G.}, \bibinfo{author}{Melosh, H.},
  \bibinfo{author}{Wünnemann, K.}, \bibinfo{year}{2011}.
\newblock \bibinfo{title}{Improvements to the $\epsilon$-$\alpha$ porous
  compaction model for simulating impacts into high-porosity solar system
  objects}.
\newblock \bibinfo{journal}{International Journal of Impact Engineering}
  \bibinfo{volume}{38}, \bibinfo{pages}{434--439}.
\newblock \URLprefix
  \url{https://www.sciencedirect.com/science/article/pii/S0734743X10001594},
  \DOIprefix\doi{https://doi.org/10.1016/j.ijimpeng.2010.10.013}.
  \bibinfo{note}{hypervelocity Impact selected papers from the 2010 Symposium}.
%Type = Inbook
\bibitem[{Collins et~al.(2019)Collins, Housen, Jutzi and Nakamura}]{Col19}
\bibinfo{author}{Collins, G.S.}, \bibinfo{author}{Housen, K.R.},
  \bibinfo{author}{Jutzi, M.}, \bibinfo{author}{Nakamura, A.M.},
  \bibinfo{year}{2019}.
\newblock \bibinfo{title}{Planetary Impact Processes in Porous Materials}.
  \bibinfo{publisher}{Springer International Publishing},
  \bibinfo{address}{Cham}.
\newblock pp. \bibinfo{pages}{103--136}.
\newblock \URLprefix \url{https://doi.org/10.1007/978-3-030-23002-9_4},
  \DOIprefix\doi{10.1007/978-3-030-23002-9_4}.
%Type = Article
\bibitem[{Collins et~al.(2004)Collins, Melosh and Ivanov}]{Collins2004}
\bibinfo{author}{Collins, G.S.}, \bibinfo{author}{Melosh, H.J.},
  \bibinfo{author}{Ivanov, B.A.}, \bibinfo{year}{2004}.
\newblock \bibinfo{title}{Modeling damage and deformation in impact
  simulations}.
\newblock \bibinfo{journal}{Meteoritics \& Planetary Science}
  \bibinfo{volume}{39}, \bibinfo{pages}{217--231}.
\newblock \URLprefix
  \url{https://onlinelibrary.wiley.com/doi/abs/10.1111/j.1945-5100.2004.tb00337.x},
  \DOIprefix\doi{https://doi.org/10.1111/j.1945-5100.2004.tb00337.x},
  \href{http://arxiv.org/abs/https://onlinelibrary.wiley.com/doi/pdf/10.1111/j.1945-5100.2004.tb00337.x}{\tt
  arXiv:https://onlinelibrary.wiley.com/doi/pdf/10.1111/j.1945-5100.2004.tb00337.x}.
%Type = Article
\bibitem[{{Davison} et~al.(2010){Davison}, {Collins} and {Ciesla}}]{Dav10}
\bibinfo{author}{{Davison}, T.M.}, \bibinfo{author}{{Collins}, G.S.},
  \bibinfo{author}{{Ciesla}, F.J.}, \bibinfo{year}{2010}.
\newblock \bibinfo{title}{{Numerical modelling of heating in porous
  planetesimal collisions}}.
\newblock \bibinfo{journal}{\icarus} \bibinfo{volume}{208},
  \bibinfo{pages}{468--481}.
\newblock \DOIprefix\doi{10.1016/j.icarus.2010.01.034}.
%Type = Article
\bibitem[{{Davison} et~al.(2011){Davison}, {Collins}, {Elbeshausen},
  {W{\"u}nnemann} and {Kearsley}}]{Dav11}
\bibinfo{author}{{Davison}, T.M.}, \bibinfo{author}{{Collins}, G.S.},
  \bibinfo{author}{{Elbeshausen}, D.}, \bibinfo{author}{{W{\"u}nnemann}, K.},
  \bibinfo{author}{{Kearsley}, A.}, \bibinfo{year}{2011}.
\newblock \bibinfo{title}{{Numerical modeling of oblique hypervelocity impacts
  on strong ductile targets}}.
\newblock \bibinfo{journal}{Meteoritics \& Planetary Science}
  \bibinfo{volume}{46}, \bibinfo{pages}{1510--1524}.
\newblock \DOIprefix\doi{10.1111/j.1945-5100.2011.01246.x}.
%Type = Inproceedings
\bibitem[{Elbeshausen and W{\"u}nnemann(2011)}]{Elbeshausen2011}
\bibinfo{author}{Elbeshausen, D.}, \bibinfo{author}{W{\"u}nnemann, K.},
  \bibinfo{year}{2011}.
\newblock \bibinfo{title}{isale-3d: A three-dimensional, multi-material,
  multi-rheology hydrocode and its applications to large-scale geodynamic
  processes}, in: \bibinfo{booktitle}{Proceedings of 11th Hypervelocity Impact
  Symposium}, \bibinfo{organization}{Fraunhofer Verlag Freiburg}. pp.
  \bibinfo{pages}{287--301}.
%Type = Article
\bibitem[{{Elbeshausen} et~al.(2009){Elbeshausen}, {W{\"u}nnemann} and
  {Collins}}]{Elbeshausen2009}
\bibinfo{author}{{Elbeshausen}, D.}, \bibinfo{author}{{W{\"u}nnemann}, K.},
  \bibinfo{author}{{Collins}, G.S.}, \bibinfo{year}{2009}.
\newblock \bibinfo{title}{{Scaling of oblique impacts in frictional targets:
  Implications for crater size and formation mechanisms}}.
\newblock \bibinfo{journal}{\icarus} \bibinfo{volume}{204},
  \bibinfo{pages}{716--731}.
\newblock \DOIprefix\doi{10.1016/j.icarus.2009.07.018}.
%Type = Article
\bibitem[{{Flynn} et~al.(1999){Flynn}, {Moore} and {Kl{\"o}ck}}]{Fly99}
\bibinfo{author}{{Flynn}, G.J.}, \bibinfo{author}{{Moore}, L.B.},
  \bibinfo{author}{{Kl{\"o}ck}, W.}, \bibinfo{year}{1999}.
\newblock \bibinfo{title}{{Density and Porosity of Stone Meteorites:
  Implications for the Density, Porosity, Cratering, and Collisional Disruption
  of Asteroids}}.
\newblock \bibinfo{journal}{\icarus} \bibinfo{volume}{142},
  \bibinfo{pages}{97--105}.
\newblock \DOIprefix\doi{10.1006/icar.1999.6210}.
%Type = Article
\bibitem[{{Genge} et~al.(2008){Genge}, {Engrand}, {Gounelle} and
  {Taylor}}]{Gen08}
\bibinfo{author}{{Genge}, M.J.}, \bibinfo{author}{{Engrand}, C.},
  \bibinfo{author}{{Gounelle}, M.}, \bibinfo{author}{{Taylor}, S.},
  \bibinfo{year}{2008}.
\newblock \bibinfo{title}{{The classification of micrometeorites}}.
\newblock \bibinfo{journal}{\maps} \bibinfo{volume}{43},
  \bibinfo{pages}{497--515}.
\newblock \DOIprefix\doi{10.1111/j.1945-5100.2008.tb00668.x}.
%Type = Article
\bibitem[{{G{\"u}ttler} et~al.(2019){G{\"u}ttler}, {Mannel}, {Rotundi},
  {Merouane}, {Fulle}, {Bockel{\'e}e-Morvan}, {Lasue}, {Levasseur-Regourd},
  {Blum}, {Naletto}, {Sierks}, {Hilchenbach}, {Tubiana}, {Capaccioni},
  {Paquette}, {Flandes}, {Moreno}, {Agarwal}, {Bodewits}, {Bertini}, {Tozzi},
  {Hornung}, {Langevin}, {Kr{\"u}ger}, {Longobardo}, {Della Corte}, {T{\'o}th},
  {Filacchione}, {Ivanovski}, {Mottola} and {Rinaldi}}]{Gut19}
\bibinfo{author}{{G{\"u}ttler}, C.}, \bibinfo{author}{{Mannel}, T.},
  \bibinfo{author}{{Rotundi}, A.}, \bibinfo{author}{{Merouane}, S.},
  \bibinfo{author}{{Fulle}, M.}, \bibinfo{author}{{Bockel{\'e}e-Morvan}, D.},
  \bibinfo{author}{{Lasue}, J.}, \bibinfo{author}{{Levasseur-Regourd}, A.C.},
  \bibinfo{author}{{Blum}, J.}, \bibinfo{author}{{Naletto}, G.},
  \bibinfo{author}{{Sierks}, H.}, \bibinfo{author}{{Hilchenbach}, M.},
  \bibinfo{author}{{Tubiana}, C.}, \bibinfo{author}{{Capaccioni}, F.},
  \bibinfo{author}{{Paquette}, J.A.}, \bibinfo{author}{{Flandes}, A.},
  \bibinfo{author}{{Moreno}, F.}, \bibinfo{author}{{Agarwal}, J.},
  \bibinfo{author}{{Bodewits}, D.}, \bibinfo{author}{{Bertini}, I.},
  \bibinfo{author}{{Tozzi}, G.P.}, \bibinfo{author}{{Hornung}, K.},
  \bibinfo{author}{{Langevin}, Y.}, \bibinfo{author}{{Kr{\"u}ger}, H.},
  \bibinfo{author}{{Longobardo}, A.}, \bibinfo{author}{{Della Corte}, V.},
  \bibinfo{author}{{T{\'o}th}, I.}, \bibinfo{author}{{Filacchione}, G.},
  \bibinfo{author}{{Ivanovski}, S.L.}, \bibinfo{author}{{Mottola}, S.},
  \bibinfo{author}{{Rinaldi}, G.}, \bibinfo{year}{2019}.
\newblock \bibinfo{title}{{Synthesis of the morphological description of
  cometary dust at comet 67P/Churyumov-Gerasimenko}}.
\newblock \bibinfo{journal}{\aap} \bibinfo{volume}{630}, \bibinfo{pages}{A24}.
\newblock \DOIprefix\doi{10.1051/0004-6361/201834751},
  \href{http://arxiv.org/abs/1902.10634}{\tt arXiv:1902.10634}.
%Type = Article
\bibitem[{{Hirt} et~al.(1974){Hirt}, {Amsden} and {Cook}}]{Hirt1974}
\bibinfo{author}{{Hirt}, C.W.}, \bibinfo{author}{{Amsden}, A.A.},
  \bibinfo{author}{{Cook}, J.L.}, \bibinfo{year}{1974}.
\newblock \bibinfo{title}{{An Arbitrary Lagrangian-Eulerian Computing Method
  for All Flow Speeds}}.
\newblock \bibinfo{journal}{Journal of Computational Physics}
  \bibinfo{volume}{14}, \bibinfo{pages}{227--253}.
\newblock \DOIprefix\doi{10.1016/0021-9991(74)90051-5}.
%Type = Article
\bibitem[{{Hyodo} and {Genda}(2020)}]{Hyo20}
\bibinfo{author}{{Hyodo}, R.}, \bibinfo{author}{{Genda}, H.},
  \bibinfo{year}{2020}.
\newblock \bibinfo{title}{{Escape and Accretion by Cratering Impacts:
  Formulation of Scaling Relations for High-speed Ejecta}}.
\newblock \bibinfo{journal}{ApJ} \bibinfo{volume}{898}, \bibinfo{pages}{30}.
\newblock \DOIprefix\doi{10.3847/1538-4357/ab9897},
  \href{http://arxiv.org/abs/2006.00732}{\tt arXiv:2006.00732}.
%Type = Article
\bibitem[{{Hyodo} and {Genda}(2021)}]{Hyo21a}
\bibinfo{author}{{Hyodo}, R.}, \bibinfo{author}{{Genda}, H.},
  \bibinfo{year}{2021}.
\newblock \bibinfo{title}{{Erosion and Accretion by Cratering Impacts on Rocky
  and Icy Bodies}}.
\newblock \bibinfo{journal}{ApJ} \bibinfo{volume}{913}, \bibinfo{pages}{77}.
\newblock \DOIprefix\doi{10.3847/1538-4357/abf6d8},
  \href{http://arxiv.org/abs/2104.04981}{\tt arXiv:2104.04981}.
%Type = Article
\bibitem[{{Hyodo} et~al.(2021){Hyodo}, {Genda} and {Brasser}}]{Hyo21b}
\bibinfo{author}{{Hyodo}, R.}, \bibinfo{author}{{Genda}, H.},
  \bibinfo{author}{{Brasser}, R.}, \bibinfo{year}{2021}.
\newblock \bibinfo{title}{{Modification of the composition and density of
  Mercury from late accretion}}.
\newblock \bibinfo{journal}{Icarus} \bibinfo{volume}{354},
  \bibinfo{pages}{114064}.
\newblock \DOIprefix\doi{10.1016/j.icarus.2020.114064},
  \href{http://arxiv.org/abs/2008.08490}{\tt arXiv:2008.08490}.
%Type = Article
\bibitem[{Hyodo et~al.(2025)Hyodo, Genda and Madeira}]{Hyo25}
\bibinfo{author}{Hyodo, R.}, \bibinfo{author}{Genda, H.},
  \bibinfo{author}{Madeira, G.}, \bibinfo{year}{2025}.
\newblock \bibinfo{title}{Pollution resistance of saturn's ring particles
  during micrometeoroid impact}.
\newblock \bibinfo{journal}{Nature Geoscience} \bibinfo{volume}{18},
  \bibinfo{pages}{44--49}.
\newblock \DOIprefix\doi{10.1038/s41561-024-01598-9}.
%Type = Article
\bibitem[{Ivanov et~al.(1997)Ivanov, Deniem and Neukum}]{Ivanov1997}
\bibinfo{author}{Ivanov, B.}, \bibinfo{author}{Deniem, D.},
  \bibinfo{author}{Neukum, G.}, \bibinfo{year}{1997}.
\newblock \bibinfo{title}{Implementation of dynamic strength models into 2d
  hydrocodes: Applications for atmospheric breakup and impact cratering}.
\newblock \bibinfo{journal}{International Journal of Impact Engineering}
  \bibinfo{volume}{20}, \bibinfo{pages}{411--430}.
\newblock \URLprefix
  \url{https://www.sciencedirect.com/science/article/pii/S0734743X97875112},
  \DOIprefix\doi{https://doi.org/10.1016/S0734-743X(97)87511-2}.
  \bibinfo{note}{hypervelocity Impact Proceedings of the 1996 Symposium}.
%Type = Article
\bibitem[{{Johnson} and {Melosh}(2014)}]{Joh14}
\bibinfo{author}{{Johnson}, B.C.}, \bibinfo{author}{{Melosh}, H.J.},
  \bibinfo{year}{2014}.
\newblock \bibinfo{title}{{Formation of melt droplets, melt fragments, and
  accretionary impact lapilli during a hypervelocity impact}}.
\newblock \bibinfo{journal}{Icarus} \bibinfo{volume}{228},
  \bibinfo{pages}{347--363}.
\newblock \DOIprefix\doi{10.1016/j.icarus.2013.10.022}.
%Type = Article
\bibitem[{{Johnson} et~al.(2015){Johnson}, {Minton}, {Melosh} and
  {Zuber}}]{Joh15}
\bibinfo{author}{{Johnson}, B.C.}, \bibinfo{author}{{Minton}, D.A.},
  \bibinfo{author}{{Melosh}, H.J.}, \bibinfo{author}{{Zuber}, M.T.},
  \bibinfo{year}{2015}.
\newblock \bibinfo{title}{{Impact jetting as the origin of chondrules}}.
\newblock \bibinfo{journal}{Nature} \bibinfo{volume}{517},
  \bibinfo{pages}{339--341}.
\newblock \DOIprefix\doi{10.1038/nature14105}.
%Type = Article
\bibitem[{Kempf et~al.(2023)Kempf, Altobelli, Schmidt, Cuzzi, Estrada and
  Srama}]{Kem23}
\bibinfo{author}{Kempf, S.}, \bibinfo{author}{Altobelli, N.},
  \bibinfo{author}{Schmidt, J.}, \bibinfo{author}{Cuzzi, J.N.},
  \bibinfo{author}{Estrada, P.R.}, \bibinfo{author}{Srama, R.},
  \bibinfo{year}{2023}.
\newblock \bibinfo{title}{Micrometeoroid infall onto saturn’s rings
  constrains their age to no more than a few hundred million years}.
\newblock \bibinfo{journal}{Science Advances} \bibinfo{volume}{9},
  \bibinfo{pages}{eadf8537}.
\newblock \URLprefix
  \url{https://www.science.org/doi/abs/10.1126/sciadv.adf8537},
  \DOIprefix\doi{10.1126/sciadv.adf8537},
  \href{http://arxiv.org/abs/https://www.science.org/doi/pdf/10.1126/sciadv.adf8537}{\tt
  arXiv:https://www.science.org/doi/pdf/10.1126/sciadv.adf8537}.
%Type = Article
\bibitem[{Kraus et~al.(2011)Kraus, Senft and Stewart}]{Kra11}
\bibinfo{author}{Kraus, R.G.}, \bibinfo{author}{Senft, L.E.},
  \bibinfo{author}{Stewart, S.T.}, \bibinfo{year}{2011}.
\newblock \bibinfo{title}{Impacts onto h2o ice: Scaling laws for melting,
  vaporization, excavation, and final crater size}.
\newblock \bibinfo{journal}{Icarus} \bibinfo{volume}{214},
  \bibinfo{pages}{724--738}.
\newblock \URLprefix
  \url{https://www.sciencedirect.com/science/article/pii/S0019103511001898},
  \DOIprefix\doi{https://doi.org/10.1016/j.icarus.2011.05.016}.
%Type = Article
\bibitem[{{Kurosawa} and {Genda}(2018)}]{Kurosawa2018}
\bibinfo{author}{{Kurosawa}, K.}, \bibinfo{author}{{Genda}, H.},
  \bibinfo{year}{2018}.
\newblock \bibinfo{title}{{Effects of Friction and Plastic Deformation in
  Shock-Comminuted Damaged Rocks on Impact Heating}}.
\newblock \bibinfo{journal}{Geophysical Research Letters} \bibinfo{volume}{45},
  \bibinfo{pages}{620--626}.
\newblock \DOIprefix\doi{10.1002/2017GL076285},
  \href{http://arxiv.org/abs/1801.01100}{\tt arXiv:1801.01100}.
%Type = Article
\bibitem[{{Levasseur-Regourd} et~al.(2018){Levasseur-Regourd}, {Agarwal},
  {Cottin}, {Engrand}, {Flynn}, {Fulle}, {Gombosi}, {Langevin}, {Lasue},
  {Mannel}, {Merouane}, {Poch}, {Thomas} and {Westphal}}]{Lev18}
\bibinfo{author}{{Levasseur-Regourd}, A.C.}, \bibinfo{author}{{Agarwal}, J.},
  \bibinfo{author}{{Cottin}, H.}, \bibinfo{author}{{Engrand}, C.},
  \bibinfo{author}{{Flynn}, G.}, \bibinfo{author}{{Fulle}, M.},
  \bibinfo{author}{{Gombosi}, T.}, \bibinfo{author}{{Langevin}, Y.},
  \bibinfo{author}{{Lasue}, J.}, \bibinfo{author}{{Mannel}, T.},
  \bibinfo{author}{{Merouane}, S.}, \bibinfo{author}{{Poch}, O.},
  \bibinfo{author}{{Thomas}, N.}, \bibinfo{author}{{Westphal}, A.},
  \bibinfo{year}{2018}.
\newblock \bibinfo{title}{{Cometary Dust}}.
\newblock \bibinfo{journal}{\ssr} \bibinfo{volume}{214}, \bibinfo{pages}{64}.
\newblock \DOIprefix\doi{10.1007/s11214-018-0496-3}.
%Type = Article
\bibitem[{{Luo} et~al.(2022){Luo}, {Zhu} and {Ding}}]{Luo2022}
\bibinfo{author}{{Luo}, X.Z.}, \bibinfo{author}{{Zhu}, M.H.},
  \bibinfo{author}{{Ding}, M.}, \bibinfo{year}{2022}.
\newblock \bibinfo{title}{{Ejecta Pattern of Oblique Impacts on the Moon From
  Numerical Simulations}}.
\newblock \bibinfo{journal}{Journal of Geophysical Research (Planets)}
  \bibinfo{volume}{127}, \bibinfo{pages}{e2022JE007333}.
\newblock \DOIprefix\doi{10.1029/2022JE007333}.
%Type = Article
\bibitem[{{Luther} et~al.(2017){Luther}, {Artemieva}, {Ivanova}, {Lorenz} and
  {W{\"u}nnemann}}]{Lut17}
\bibinfo{author}{{Luther}, R.}, \bibinfo{author}{{Artemieva}, N.},
  \bibinfo{author}{{Ivanova}, M.}, \bibinfo{author}{{Lorenz}, C.},
  \bibinfo{author}{{W{\"u}nnemann}, K.}, \bibinfo{year}{2017}.
\newblock \bibinfo{title}{{Snow carrots after the Chelyabinsk event and model
  implications for highly porous solar system objects}}.
\newblock \bibinfo{journal}{Meteoritics \& Planetary Science}
  \bibinfo{volume}{52}, \bibinfo{pages}{979--999}.
\newblock \DOIprefix\doi{10.1111/maps.12831}.
%Type = Article
\bibitem[{{Manske} et~al.(2022){Manske}, {W{\"u}nnemann} and
  {Kurosawa}}]{Man22}
\bibinfo{author}{{Manske}, L.}, \bibinfo{author}{{W{\"u}nnemann}, K.},
  \bibinfo{author}{{Kurosawa}, K.}, \bibinfo{year}{2022}.
\newblock \bibinfo{title}{{Quantification of Impact-Induced Melt Production in
  Numerical Modeling Revisited}}.
\newblock \bibinfo{journal}{Journal of Geophysical Research (Planets)}
  \bibinfo{volume}{127}, \bibinfo{pages}{e2022JE007426}.
\newblock \DOIprefix\doi{10.1029/2022JE007426}.
%Type = Article
\bibitem[{{Melosh}(1984)}]{Mel84}
\bibinfo{author}{{Melosh}, H.J.}, \bibinfo{year}{1984}.
\newblock \bibinfo{title}{{Impact ejection, spallation, and the origin of
  meteorites}}.
\newblock \bibinfo{journal}{Icarus} \bibinfo{volume}{59},
  \bibinfo{pages}{234--260}.
\newblock \DOIprefix\doi{10.1016/0019-1035(84)90026-5}.
%Type = Article
\bibitem[{MELOSH(2007)}]{Mel07}
\bibinfo{author}{MELOSH, H.J.}, \bibinfo{year}{2007}.
\newblock \bibinfo{title}{A hydrocode equation of state for sio2}.
\newblock \bibinfo{journal}{Meteoritics \& Planetary Science}
  \bibinfo{volume}{42}, \bibinfo{pages}{2079--2098}.
\newblock \URLprefix
  \url{https://onlinelibrary.wiley.com/doi/abs/10.1111/j.1945-5100.2007.tb01009.x},
  \DOIprefix\doi{https://doi.org/10.1111/j.1945-5100.2007.tb01009.x},
  \href{http://arxiv.org/abs/https://onlinelibrary.wiley.com/doi/pdf/10.1111/j.1945-5100.2007.tb01009.x}{\tt
  arXiv:https://onlinelibrary.wiley.com/doi/pdf/10.1111/j.1945-5100.2007.tb01009.x}.
%Type = Article
\bibitem[{Melosh et~al.(1992)Melosh, Ryan and Asphaug}]{Mel92}
\bibinfo{author}{Melosh, H.J.}, \bibinfo{author}{Ryan, E.V.},
  \bibinfo{author}{Asphaug, E.}, \bibinfo{year}{1992}.
\newblock \bibinfo{title}{Dynamic fragmentation in impacts: Hydrocode
  simulation of laboratory impacts}.
\newblock \bibinfo{journal}{Journal of Geophysical Research: Planets}
  \bibinfo{volume}{97}, \bibinfo{pages}{14735--14759}.
\newblock \URLprefix
  \url{https://agupubs.onlinelibrary.wiley.com/doi/abs/10.1029/92JE01632},
  \DOIprefix\doi{https://doi.org/10.1029/92JE01632},
  \href{http://arxiv.org/abs/https://agupubs.onlinelibrary.wiley.com/doi/pdf/10.1029/92JE01632}{\tt
  arXiv:https://agupubs.onlinelibrary.wiley.com/doi/pdf/10.1029/92JE01632}.
%Type = Article
\bibitem[{{Okamoto} et~al.(2013){Okamoto}, {Nakamura}, {Hasegawa}, {Kurosawa},
  {Ikezaki} and {Tsuchiyama}}]{Oka13}
\bibinfo{author}{{Okamoto}, T.}, \bibinfo{author}{{Nakamura}, A.M.},
  \bibinfo{author}{{Hasegawa}, S.}, \bibinfo{author}{{Kurosawa}, K.},
  \bibinfo{author}{{Ikezaki}, K.}, \bibinfo{author}{{Tsuchiyama}, A.},
  \bibinfo{year}{2013}.
\newblock \bibinfo{title}{{Impact experiments of exotic dust grain capture by
  highly porous primitive bodies}}.
\newblock \bibinfo{journal}{Icarus} \bibinfo{volume}{224},
  \bibinfo{pages}{209--217}.
\newblock \DOIprefix\doi{10.1016/j.icarus.2013.02.023}.
%Type = Article
\bibitem[{Pierazzo et~al.(1997)Pierazzo, Vickery and Melosh}]{Pie97}
\bibinfo{author}{Pierazzo, E.}, \bibinfo{author}{Vickery, A.},
  \bibinfo{author}{Melosh, H.}, \bibinfo{year}{1997}.
\newblock \bibinfo{title}{A reevaluation of impact melt production}.
\newblock \bibinfo{journal}{Icarus} \bibinfo{volume}{127},
  \bibinfo{pages}{408--423}.
\newblock \URLprefix
  \url{https://www.sciencedirect.com/science/article/pii/S0019103597957134},
  \DOIprefix\doi{https://doi.org/10.1006/icar.1997.5713}.
%Type = Article
\bibitem[{{Pokorn{\'y}} et~al.(2017){Pokorn{\'y}}, {Sarantos} and
  {Janches}}]{Pok17}
\bibinfo{author}{{Pokorn{\'y}}, P.}, \bibinfo{author}{{Sarantos}, M.},
  \bibinfo{author}{{Janches}, D.}, \bibinfo{year}{2017}.
\newblock \bibinfo{title}{{Reconciling the Dawn-Dusk Asymmetry in
  Mercury{\textquoteright}s Exosphere with the Micrometeoroid Impact
  Directionality}}.
\newblock \bibinfo{journal}{The Astrophysical Journal Letters}
  \bibinfo{volume}{842}, \bibinfo{pages}{L17}.
\newblock \DOIprefix\doi{10.3847/2041-8213/aa775d},
  \href{http://arxiv.org/abs/1706.01461}{\tt arXiv:1706.01461}.
%Type = Article
\bibitem[{{Senft} and {Stewart}(2008)}]{Sen08}
\bibinfo{author}{{Senft}, L.E.}, \bibinfo{author}{{Stewart}, S.T.},
  \bibinfo{year}{2008}.
\newblock \bibinfo{title}{{Impact crater formation in icy layered terrains on
  Mars}}.
\newblock \bibinfo{journal}{Meteoritics and Planetary Science}
  \bibinfo{volume}{43}, \bibinfo{pages}{1993--2013}.
\newblock \DOIprefix\doi{10.1111/j.1945-5100.2008.tb00657.x}.
%Type = Article
\bibitem[{{Silber} and {Johnson}(2017)}]{Sil17}
\bibinfo{author}{{Silber}, E.A.}, \bibinfo{author}{{Johnson}, B.C.},
  \bibinfo{year}{2017}.
\newblock \bibinfo{title}{{Impact Crater Morphology and the Structure of
  Europa's Ice Shell}}.
\newblock \bibinfo{journal}{Journal of Geophysical Research (Planets)}
  \bibinfo{volume}{122}, \bibinfo{pages}{2685--2701}.
\newblock \DOIprefix\doi{10.1002/2017JE005456},
  \href{http://arxiv.org/abs/1711.08997}{\tt arXiv:1711.08997}.
%Type = Article
\bibitem[{{Stewart} and {Ahrens}(2005)}]{Ste05}
\bibinfo{author}{{Stewart}, S.T.}, \bibinfo{author}{{Ahrens}, T.J.},
  \bibinfo{year}{2005}.
\newblock \bibinfo{title}{{Shock properties of H$_{2}$O ice}}.
\newblock \bibinfo{journal}{Journal of Geophysical Research (Planets)}
  \bibinfo{volume}{110}, \bibinfo{pages}{E03005}.
\newblock \DOIprefix\doi{10.1029/2004JE002305}.
%Type = Article
\bibitem[{Stewart et~al.(2008)Stewart, Seifter and Obst}]{Ste08}
\bibinfo{author}{Stewart, S.T.}, \bibinfo{author}{Seifter, A.},
  \bibinfo{author}{Obst, A.W.}, \bibinfo{year}{2008}.
\newblock \bibinfo{title}{Shocked h2o ice: Thermal emission measurements and
  the criteria for phase changes during impact events}.
\newblock \bibinfo{journal}{Geophysical Research Letters} \bibinfo{volume}{35}.
\newblock \URLprefix
  \url{https://agupubs.onlinelibrary.wiley.com/doi/abs/10.1029/2008GL035947},
  \DOIprefix\doi{https://doi.org/10.1029/2008GL035947},
  \href{http://arxiv.org/abs/https://agupubs.onlinelibrary.wiley.com/doi/pdf/10.1029/2008GL035947}{\tt
  arXiv:https://agupubs.onlinelibrary.wiley.com/doi/pdf/10.1029/2008GL035947}.
%Type = Article
\bibitem[{{Svetsov} and {Shuvalov}(2015)}]{Svetsov2015}
\bibinfo{author}{{Svetsov}, V.V.}, \bibinfo{author}{{Shuvalov}, V.V.},
  \bibinfo{year}{2015}.
\newblock \bibinfo{title}{{Water delivery to the Moon by asteroidal and
  cometary impacts}}.
\newblock \bibinfo{journal}{Planetary and Space Science} \bibinfo{volume}{117},
  \bibinfo{pages}{444--452}.
\newblock \DOIprefix\doi{10.1016/j.pss.2015.09.011}.
%Type = Misc
\bibitem[{{Thompson} and {Lauson}(1972)}]{Thompson1972}
\bibinfo{author}{{Thompson}, S.L.}, \bibinfo{author}{{Lauson}, H.S.},
  \bibinfo{year}{1972}.
\newblock \bibinfo{title}{{Improvements in the Chart D Radiation-Hydrodynamic
  Code. III: Revised Analytic Equations of State}}.
\newblock \bibinfo{howpublished}{Albuquerque, New Mexico: Sandia National
  Laboratory. 1972. Technical Report SC-RR-71-0714}.
%Type = Techreport
\bibitem[{Tillotson(1962)}]{tillotson1962}
\bibinfo{author}{Tillotson, J.H.}, \bibinfo{year}{1962}.
\newblock \bibinfo{title}{Metallic equations of state for hypervelocity
  impact}.
\newblock \bibinfo{type}{Technical Report}.
%Type = Article
\bibitem[{{Turtle} and {Pierazzo}(2001)}]{Turtle2001}
\bibinfo{author}{{Turtle}, E.P.}, \bibinfo{author}{{Pierazzo}, E.},
  \bibinfo{year}{2001}.
\newblock \bibinfo{title}{{Thickness of a Europan Ice Shell from Impact Crater
  Simulations}}.
\newblock \bibinfo{journal}{Science} \bibinfo{volume}{294},
  \bibinfo{pages}{1326--1328}.
\newblock \DOIprefix\doi{10.1126/science.1062492}.
%Type = Article
\bibitem[{{Wakita} et~al.(2019){Wakita}, {Genda}, {Kurosawa} and
  {Davison}}]{Wakita2019}
\bibinfo{author}{{Wakita}, S.}, \bibinfo{author}{{Genda}, H.},
  \bibinfo{author}{{Kurosawa}, K.}, \bibinfo{author}{{Davison}, T.M.},
  \bibinfo{year}{2019}.
\newblock \bibinfo{title}{{Enhancement of Impact Heating in
  Pressure-Strengthened Rocks in Oblique Impacts}}.
\newblock \bibinfo{journal}{Geophysical Research Letters} \bibinfo{volume}{46},
  \bibinfo{pages}{13,678--13,686}.
\newblock \DOIprefix\doi{10.1029/2019GL085174},
  \href{http://arxiv.org/abs/1912.00371}{\tt arXiv:1912.00371}.
%Type = Article
\bibitem[{{Wakita} et~al.(2022){Wakita}, {Genda}, {Kurosawa}, {Davison} and
  {Johnson}}]{Wakita2022}
\bibinfo{author}{{Wakita}, S.}, \bibinfo{author}{{Genda}, H.},
  \bibinfo{author}{{Kurosawa}, K.}, \bibinfo{author}{{Davison}, T.M.},
  \bibinfo{author}{{Johnson}, B.C.}, \bibinfo{year}{2022}.
\newblock \bibinfo{title}{{Effect of Impact Velocity and Angle on Deformational
  Heating and Postimpact Temperature}}.
\newblock \bibinfo{journal}{Journal of Geophysical Research (Planets)}
  \bibinfo{volume}{127}, \bibinfo{pages}{e07266}.
\newblock \DOIprefix\doi{10.1029/2022JE007266},
  \href{http://arxiv.org/abs/2208.07630}{\tt arXiv:2208.07630}.
%Type = Article
\bibitem[{{Wakita} et~al.(2021){Wakita}, {Johnson}, {Denton} and
  {Davison}}]{Wakita2021}
\bibinfo{author}{{Wakita}, S.}, \bibinfo{author}{{Johnson}, B.C.},
  \bibinfo{author}{{Denton}, C.A.}, \bibinfo{author}{{Davison}, T.M.},
  \bibinfo{year}{2021}.
\newblock \bibinfo{title}{{Jetting during oblique impacts of spherical
  impactors}}.
\newblock \bibinfo{journal}{Icarus} \bibinfo{volume}{360},
  \bibinfo{pages}{114365}.
\newblock \DOIprefix\doi{10.1016/j.icarus.2021.114365},
  \href{http://arxiv.org/abs/2102.02303}{\tt arXiv:2102.02303}.
%Type = Article
\bibitem[{{Wakita} et~al.(2017){Wakita}, {Matsumoto}, {Oshino} and
  {Hasegawa}}]{Wakita2017}
\bibinfo{author}{{Wakita}, S.}, \bibinfo{author}{{Matsumoto}, Y.},
  \bibinfo{author}{{Oshino}, S.}, \bibinfo{author}{{Hasegawa}, Y.},
  \bibinfo{year}{2017}.
\newblock \bibinfo{title}{{Planetesimal Collisions as a Chondrule Forming
  Event}}.
\newblock \bibinfo{journal}{The Astrophysical Journal} \bibinfo{volume}{834},
  \bibinfo{pages}{125}.
\newblock \DOIprefix\doi{10.3847/1538-4357/834/2/125},
  \href{http://arxiv.org/abs/1611.05511}{\tt arXiv:1611.05511}.
%Type = Article
\bibitem[{Wünnemann et~al.(2006)Wünnemann, Collins and
  Melosh}]{Wunnemann2006}
\bibinfo{author}{Wünnemann, K.}, \bibinfo{author}{Collins, G.},
  \bibinfo{author}{Melosh, H.}, \bibinfo{year}{2006}.
\newblock \bibinfo{title}{A strain-based porosity model for use in hydrocode
  simulations of impacts and implications for transient crater growth in porous
  targets}.
\newblock \bibinfo{journal}{Icarus} \bibinfo{volume}{180},
  \bibinfo{pages}{514--527}.
\newblock \URLprefix
  \url{https://www.sciencedirect.com/science/article/pii/S0019103505004124},
  \DOIprefix\doi{https://doi.org/10.1016/j.icarus.2005.10.013}.
%Type = Article
\bibitem[{{Zhang} et~al.(2017){Zhang}, {Hayes}, {Janssen}, {Nicholson},
  {Cuzzi}, {de Pater} and {Dunn}}]{Zha17a}
\bibinfo{author}{{Zhang}, Z.}, \bibinfo{author}{{Hayes}, A.G.},
  \bibinfo{author}{{Janssen}, M.A.}, \bibinfo{author}{{Nicholson}, P.D.},
  \bibinfo{author}{{Cuzzi}, J.N.}, \bibinfo{author}{{de Pater}, I.},
  \bibinfo{author}{{Dunn}, D.E.}, \bibinfo{year}{2017}.
\newblock \bibinfo{title}{{Exposure age of Saturn's A and B rings, and the
  Cassini Division as suggested by their non-icy material content}}.
\newblock \bibinfo{journal}{Icarus} \bibinfo{volume}{294},
  \bibinfo{pages}{14--42}.
\newblock \DOIprefix\doi{10.1016/j.icarus.2017.04.008}.

\end{thebibliography}



\end{document}